\documentclass[12pt,aps,showpacs,eqsecnum,nofootinbib,floatfix]{revtex4}
\usepackage{bbding}
\usepackage{pifont}
\usepackage{subfigure}
\usepackage{epsfig}
\usepackage{amsmath}
\usepackage{amsfonts}
\usepackage{amssymb}
\usepackage{color}
\usepackage{graphicx}
\usepackage{mathrsfs}
\usepackage{multirow}

\def\be{\begin{equation}}
\def\ee{\end{equation}}
\def\ba{\begin{eqnarray}}
\def\ea{\end{eqnarray}}
\def\no{\nonumber}

\newcommand{\omits}[1]{}

\def\NPB{{Nucl. Phys.}~{\bf B}}

\def\CMP{{Commun. Math. Phys. }}

\def\IJMPA{{Int. J. Mod. Phys.}~{\bf A}}

\begin{document}

\title{Thermodynamics of phase transition in higher dimensional
Reissner-Nordstr\"{o}m-de Sitter black hole }

\author{Li-Chun Zhang, Meng-Sen Ma, Hui-Hua Zhao, Ren Zhao\footnote{Email: zhao2969@sina.com}}

\medskip

\affiliation{Institute of Theoretical Physics, Shanxi Datong University,
Datong 037009, China}
\affiliation{Department of Physics, Shanxi Datong University,
Datong 037009, China}

\begin{abstract}

It is well known that there are black hole and the cosmological
horizons for the Reissner-Nordstr\"{o}m-de Sitter spacetime.
Although the thermodynamic quantities on the horizons are not
irrelevant, they satisfy the laws of black hole thermodynamics
respectively. In this paper by considering the relations between the
two horizons we give the effective thermodynamic quantities in
$(n+2)$-dimensional Reissner-Nordstr\"{o}m-de Sitter spacetime. The
thermodynamic properties of these effective quantities are analyzed,
moreover, the critical temperature, critical pressure and critical
volume are obtained. We carry out an analytical check of Ehrenfest
equations and prove that both Ehrenfest equations are satisfied. So
the spacetime undergoes a second order phase transition at the
critical point. This result is consistent with the nature of
liquid--gas phase transition at the critical point, hence deepening
the understanding of the analogy of charged dS spacetime and
liquid--gas systems.


\end{abstract}


\maketitle

\bigskip

\section{Introduction}

Black hole physics, specially black hole thermodynamics, refer to the
theories of gravity, statistical mechanics, particle physics and field
theory and so on. This makes it receive a lot of attention\cite{JDB1,JDB2,JDB3,BCH,Hawking1,Hawking2}. Although
the statistical explanation of the thermodynamic states of black holes is
still lack, the properties of black hole thermodynamics are worth
studying deeply, such as Hawking-Page phase transition\cite{Hawking3}, the critical
phenomena of black holes. More interestingly, recently through the study of
RN-AdS black hole it is shown that there exists similar phase transition to
the one in the van der Waals-Maxwell vapor-liquid system\cite{Chamblin1,Chamblin2}.

Motivated by the AdS/CFT correspondence\cite{Maldacena}, where the transitions have
been related with the holographic superconductivity\cite{Gubser,Hartnoll}, the subject
that the phase transitions of black holes in asymptotically anti de-Sitter
(AdS) spacetime has received considerable attention\cite{Sahay1,Sahay2,Sahay3,Banerjee,Kastor}. The
underlying microscopic statistical interaction of the black holes is also
expected to be understood via the study of the gauge theory living on the
boundary in the gauge/gravity duality.

Recently, by considering the cosmological constant correspond to
pressure in general thermodynamic system, namely \be
P=-\frac{1}{8\pi }\Lambda =\frac{3}{8\pi }\frac{1}{l^2}, \ee the
thermodynamic volumes in AdS and dS spacetime are
obtained\cite{Dolan1,SG,BPDolan,Cvetic,Dolan2,NA1,Dolan3}. The
studies on phase transition of black holes have aroused great
interest\cite{Banerjee,DCZ,RBM,LYX,Cai1,Liu,Hendi,Spallucci,Belhaj,Zhao1,MB,LWB,JXM,Banerjee1,Banerjee3,Banerjee4,Banerjee5,NA,
Ma}. Connecting the thermodynamic quantities of AdS black holes to
$(P\sim V)$in the ordinary thermodynamic system, the critical
behaviors of black holes can be analyzed and the phase diagram like
van der Waals vapor-liquid system can be obtained. This helps to
further understand the black hole entropy, temperature, heat
capacities, etc. It also has a very important significance in
completing the geometric theory of black hole thermodynamics.

As is well known, there are black hole horizon and the cosmological horizon
in the appropriate range of parameters for de Sitter spacetime. The both
horizons have thermal radiation, but with different temperatures. The
thermodynamic quantities on the both horizons satisfy the first law of
thermodynamics, and the corresponding entropy fulfill the area formula\cite{Dolan2,Cai2,Sekiwa}.
In recent years, the researches on the thermodynamic properties of de
Sitter spacetime have drawn many attentions\cite{Banerjee,Dolan2,Cai2,Sekiwa,Urano,Zhang,Myung,Kim}. In the inflation epoch of early universe, the universe is a quasi-de Sitter spacetime. The
cosmological constant introduced in de Sitter space may come from the vacuum
energy, which is also a kind of energy. If the cosmological constant is the
dark energy, the universe will evolve to a new de Sitter phase. To depict
the whole history of evolution of the universe, we should have some
knowledge on the classical and quantum properties of de Sitter space\cite{Dolan2,Sekiwa,Cai3,SB}.

Firstly, we expect the thermodynamic entropy to satisfy the Nernst
theorem\cite{Urano,Myung,Cai4}. At present a satisfactory explanation to the problem
that the thermodynamic entropy of the horizon of the extreme de Sitter
spacetime do not fulfill the Nernst theorem is still lacked. Secondly, when
considering the correlation between the black hole horizon and the
cosmological horizon whether the thermodynamic quantities in de Sitter
spacetime still have the phase transition and critical behavior like in AdS
black holes. Thus it is worthy of our deep investigation and reflection to
establish a consistent thermodynamics in de Sitter spacetime.

Because the thermodynamic quantities on the black hole horizon and
the cosmological one in de Sitter spacetime are the functions of
mass $M$, electric charge $Q$ and cosmological constant $\Lambda $.
The quantities are not independent each other. Considering the
relation between the thermodynamic quantities on the two horizons is
very important for studying the thermodynamic properties of de
Sitter spacetime. Based on the relation we give the effective
temperature and pressure of ($n+2)$-dimensional
Reissner-Nordstr\"{o}m-de Sitter(RN-dS) spacetime and analyze the
critical behavior of the effective thermodynamic quantities.

The paper is arranged as follows: in the Sec.II we introduce the
($n+2)$-dimensional Reissner-Nordstr\"{o}mde Sitter(RN-dS) spacetime,
and give the two horizons and corresponding thermodynamic
quantities; In Sec.III by considering the relations between the two
horizons we obtain the effective temperature and the equivalent
pressure; In Sec. IV the critical phenomena of effective
thermodynamic quantities is discussed; To investigate the nature of
the phase transition at the critical point, we will introduce the
classical Ehrenfest scheme and carry out an analytical check of both
equations in Section V. Finally we discuss and summarize our results
in Sec.VI. (we use the units $G_{n+1} =\hbar =k_B =c=1)$

\section{RN-dS spacetime}

The line element of the ($n+2)$ dimensional RN-dS black holes is given
by\cite{Cai2}
\begin{equation}
\label{eq1}
ds^2=-f(r)dt^2+f^{-1}dr^2+r^2d\Omega _n^2 ,
\end{equation}
where
\ba
\label{eq2}
&&f(r)=1-\frac{\omega _n M}{r^{n-1}}+\frac{n\omega _n^2
Q^2}{8(n-1)r^{2n-2}}-\frac{2\Lambda }{n(n+1)}r^2,\no\\
&&\omega _n =\frac{16\pi }{nVol(S^n)}=\frac{8\Gamma
(\textstyle{{n+1} \over 2})}{n\pi^{\textstyle{{n-1} \over 2}}}, \ea
in which $Q$ is the electric/magnetic charge of Maxwell field,
$\Lambda >0$ -de Sitter£¬$\Lambda <0$ -anti-de Sitter. For general
$M$ and $Q$, $l=\sqrt {\frac{n(n+1)}{2\Lambda }} $ is the curvature
radius of dS space, $Vol(S^n)$ denotes the volume of a unit
$n$-sphere $d\Omega _n^2 $, the equation $f(r)=0$ may have four real
roots when the parameters $M, Q, l$ satisfy a
condition\cite{realroots}. Three of them are positive: the largest
one is the cosmological horizon (CEH) $r=r_c $, the smallest is the
inner (Cauchy) horizon of black hole, the middle one is the outer
horizon (BEH) $r=r_+ $ of black hole.
\[
d\Omega _n^2 =d\chi _2^2 +\sin ^2\chi _2^2 d\chi _3^2 +\cdots
\prod\limits_{i=2}^n {\sin _i^2 } d\chi _{n+1}^2.
\]
The equations $f(r_+ )=0$ and $f(r_c )=0$ are rearranged to
\ba
Q^2&=&\frac{8(n-1)(r_+ r_c )^{n-1}}{n\omega _n^2 }\left( {1-\frac{2(r_c^{n+1}
-r_+^{n+1} )}{n(n+1)(r_c^{n-1} -r_+^{n-1} )}\Lambda } \right),\no\\
\frac{2\Lambda }{n(n+1)}&=&\left( {1-\frac{n\omega _n^2 Q^2}{8(n-1)(r_c r_+
)^{n-1}}} \right)\left( {\frac{r_c^{n-1} -r_+^{n-1} }{r_c^{n+1} -r_+^{n+1}
}} \right),\label{eq3}\\
\label{eq4}\omega _n M&=&(r_c^{n-1} +r_+^{n-1} )-\frac{2\Lambda (r_c^{2n} -r_+^{2n}
)}{n(n+1)(r_c^{n-1} -r_+^{n-1} )}\no\\
&=&(r_c^{n-1} +r_+^{n-1} )-\frac{(r_c^{2n} -r_+^{2n} )}{(r_c^{n+1} -r_+^{n+1}
)}\left( {1-\frac{n\omega _n^2 Q^2}{8(n-1)(r_c r_+ )^{n-1}}} \right)\no\\
&=&\frac{(r_c r_+ )^{n-1}(r_c^2 -r_+^2 )}{(r_c^{n+1} -r_+^{n+1}
)}+\frac{(r_c^{2n} -r_+^{2n} )}{(r_c^{n+1} -r_+^{n+1} )}\frac{n\omega _n^2
Q^2}{8(n-1)(r_c r_+ )^{n-1}}.
\ea
The surface gravities at the black hole horizon and the cosmological horizon
are respectively
\begin{equation}
\label{eq5}
\kappa _+ =\frac{1}{2}\left. {\frac{df(r)}{dr}} \right|_{r=r_+ }
=\frac{1}{2r_+ }\left( {(n-1)-\frac{2\Lambda }{n}r_+^2 -\frac{n\omega _n^2
Q^2}{8r_+^{2n-2} }} \right),
\end{equation}
\begin{equation}
\label{eq6}
\kappa _c =\frac{1}{2}\left. {\frac{df(r)}{dr}} \right|_{r=r_c }
=\frac{1}{2r_c }\left( {(n-1)-\frac{2\Lambda }{n}r_c^2 -\frac{n\omega _n^2
Q^2}{8r_c^{2n-2} }} \right).
\end{equation}
The thermodynamic quantities corresponding to the two horizons satisfy the
first law of thermodynamics respectively\cite{Dolan2,Sekiwa,Gibbons1,Gibbons2}
\ba
\label{eq7}
\delta M&=&\frac{\kappa _+ }{2\pi }\delta S_+ +\Phi _+ \delta Q+V_+ \delta P,\\
\label{eq8} \delta M&=&\frac{\kappa _c }{2\pi }\delta S_c +\Phi _c
\delta Q+V_c \delta P, \ea

where \[ S_+ =\frac{r_+^n Vol(S^n)}{4G}, \quad
V_+=\frac{Vol(S^n)}{n+1}r_+^{n+1},\quad \Phi _+
=\frac{n}{4(n-1)}\frac{\omega _n Q}{r_+^{n-1} }, \quad S_c
=\frac{r_c^n Vol(S^n)}{4G} , \]
 \begin{equation}
\label{eq8}V_c =\frac{Vol(S^n)}{n+1}r_c^{n+1},\quad \Phi
_c=-\frac{n}{4(n-1)}\frac{\omega _n Q}{r_c^{n-1} },\quad
P=-\frac{n\Lambda }{16\pi }. \end{equation}

\section{Thermodynamic quantity of RN-dS spacetime}

In the above section, we have obtained thermodynamic quantities without considering
the relationship between the black hole horizon and the cosmological
horizon. Because there are three variables $M$, $Q$ and $\Lambda $ in the
spacetime, the thermodynamic quantities corresponding to the black hole
horizon and the cosmological horizon are function with respect to $M$, $Q$
and $\Lambda $. The thermodynamic quantities corresponding to the black hole
horizon are related to the ones corresponding to the cosmological horizon.
When the thermodynamic property of charged de Sitter spacetime is studied,
we must consider the relationship with the two horizon, Recently, by
studying Hawking radiation of de Sitter spacetime, in \cite{Zhao2,Zhao3} it is obtained
that the outgoing rate of the charged de Sitter spacetime which radiates
particles with energy $\omega $ is
\begin{equation}
\label{eq9}
\Gamma =e^{\Delta S_+ +\Delta S_c },
\end{equation}
where $\Delta S_+ $ and $\Delta S_c $ are Bekenstein-Hawking entropy
difference corresponding to the black hole horizon and the cosmological
horizon after the charged de Sitter spacetime radiates particles with energy
$\omega $. Therefore, the thermodynamic entropy of the charged de Sitter
spacetime is the sum of the black hole horizon entropy and the cosmological
horizon entropy
\begin{equation}
\label{eq10}
S=S_+ +S_c .
\end{equation}
Substitute Eqs.(\ref{eq7}) and (2.8) into Eq.(\ref{eq10}), we can
get
\begin{equation}
\label{eq11}
dS=2\pi \left( {\frac{1}{\kappa _+ }+\frac{1}{\kappa _c }}
\right)dM-\frac{n\pi \omega _n Q}{2(n-1)}\left( {\frac{1}{r_+^{n-1} \kappa
_+ }+\frac{1}{r_c^{n-1} \kappa _c }} \right)dQ+\frac{Vol(S^n)}{4(n+1)}\left(
{\frac{r_+^{n+1} }{\kappa _+ }+\frac{r_c^{n+1} }{\kappa _c }}
\right)d\Lambda .
\end{equation}
For simplicity, we set a fixed $Q$ . In this case the above equation turns into
\begin{equation}
\label{eq12}
dS=2\pi \left( {\frac{1}{\kappa _+ }+\frac{1}{\kappa _c }}
\right)dM+\frac{Vol(S^n)}{4(n+1)}\left( {\frac{r_+^{n+1} }{\kappa _+
}+\frac{r_c^{n+1} }{\kappa _c }} \right)d\Lambda .
\end{equation}
From Eqs.(\ref{eq3}), (\ref{eq4}), (\ref{eq5}) and (\ref{eq6}) one can obtain
\[
dr_c =\frac{1}{2\kappa _c }\left( {\frac{\omega _n dM}{r_c^{n-1}
}-\frac{n\omega _n^2 QdQ}{4(n-1)r_c^{2n-2} }+\frac{2r_c^2 }{n(n+1)}d\Lambda
} \right),
\]
\begin{equation}
\label{eq13}
dr_+ =\frac{1}{2\kappa _+ }\left( {\frac{\omega _n dM}{r_+^{n-1}
}-\frac{n\omega _n^2 QdQ}{4(n-1)r_+^{2n-2} }+\frac{2r_+^2 }{n(n+1)}d\Lambda
} \right).
\end{equation}
Recently, in \cite{Dolan2} the thermodynamic volume of higher dimensional RN-dS black
hole is given as
\begin{equation}
\label{eq14}
V=\frac{Vol(S^n)}{n+1}\left( {r_c^{n+1} -r_+^{n+1} } \right).
\end{equation}
Substitute Eq. (\ref{eq13}) into Eq. (\ref{eq14}), one can get
\[
dV=\frac{\omega _n Vol(S^n)}{2}\left( {\frac{r_c }{\kappa _c }-\frac{r_+
}{\kappa _+ }} \right)dM+\frac{Vol(S^n)}{n(n+1)}\left( {\frac{r_c^{n+2}
}{\kappa _c }-\frac{r_+^{n+2} }{\kappa _+ }} \right)d\Lambda
\]
\begin{equation}
\label{eq15}
-\frac{2\pi \omega _n Q}{(n-1)}\left( {\frac{1}{r_c^{n-2} \kappa _c
}-\frac{1}{r_+^{n-2} \kappa _+ }} \right)dQ.
\end{equation}
Substituting Eq. (\ref{eq15}) into Eq. (\ref{eq11}), one can obtain
the thermodynamic equation for the thermodynamic quantities of
higher dimensional RN-dS black hole\cite{Urano}.
\begin{equation}
\label{eq16}
dM=T_{eff} dS-P_{eff} dV+\varphi _{eff} dQ,
\end{equation}
where
\begin{equation}
\label{eq19}
T_{eff} =\frac{B_1 }{4\pi r_c x(1+x)}
+\frac{n\omega _n^2 Q^2B_2 }{32\pi r_c^{2n-1} (n-1)x^{2n-1}(1+x)},
\end{equation}
\begin{equation}
\label{eq20}
P_{eff} =\frac{B_3 }{16\pi r_c^2 x(1+x)}
+\frac{n^2\omega _n^2 Q^2B_4 }{16\times 8\pi (n-1)r_c^{2n} x^{2n-1}(1+x)},
\end{equation}
\begin{equation}
\label{eq21}
\varphi _{eff} =\frac{n\omega _n Q}{4(n-1)}\frac{(1-x^{2n})}{r_c^{n-1}
x^{n-1}(1-x^{n+1})},
\end{equation}
and
\[
(1-x^{n+1})B_1 =n(1-x^2)(1+x^{n+1})-(1+x^2)(1-x^{n+1}),
\]
\[
(1-x^{n+1})B_2 =(1-x^{2n})(1+x^{n+1})-n(1+x^{2n})(1-x^{n+1}),
\]
\[
(1-x^{n+1})^2B_3
=n[n(1-x^2)(1+x^{2n+1})-(1+x^2)(1-x^{2n+1})+2x^{n+1}(1-x)],
\]
\begin{equation}
\label{eq22}
(1-x^{n+1})^2B_4 =2nx^{n+1}(1-x^{2n-1})-(n-1)(1-x^{4n+1})-(n+1)x^{2n}(1-x).
\end{equation}
$x:=r_+ /r_c $ and $0<x<1$.\footnote{It should be noted that the parameters $M, Q, l$ should satisfy a condition to guarantee four real roots to exist. Under the condition, the $x$ here cannot tend to zero. }

The thermodynamic quantities defined in (\ref{eq19}), (\ref{eq20}) and (\ref{eq21}) meet the thermodynamic equation (\ref{eq16}). When $x\to 1$, namely
the both horizons coincide,
\[
P_{eff} \to 0,
\]
\begin{equation}
\label{eq23}
T_{eff} \to \frac{(n-1)}{4\pi r_c (n+1)}-\frac{n^2\omega _n^2 Q^2}{32\pi
r_c^{2n-1} (n+1)}
=\frac{1}{4\pi r_c (n+1)}\left( {2r_c^2 \Lambda -(n-1)^2} \right).
\end{equation}
In this case, from Eqs. (\ref{eq3}) and (\ref{eq4}), we have
\begin{equation}
\label{eq24}
Q^2=\frac{8(n-1)r_c^{2(n-1)} }{n\omega _n^2 }\left( {1-\frac{2r_c^2
}{n(n-1)}\Lambda } \right),
\quad
\omega _n M=2r_c^{n-1} -\frac{4r_c^{n+1} \Lambda }{(n+1)(n-1)},
\end{equation}
Due to $Q^2\ge 0$,$M\ge 0$, thus $\frac{2r_c^2 \Lambda }{n(n-1)}\le 1$. When
$M^2=Q^2$, from Eq. (\ref{eq24}) one can obtain
\begin{equation}
\label{eq25}
2r_c^2 \Lambda =\frac{(n+1)(n-1)}{n^2}\left( {1+\sqrt {1+n^3(n-2)} }
\right).
\end{equation}
Therefore, when $M^2\ge Q^2$, there should be the relation $2r_c^2 \Lambda
\ge \frac{(n+1)(n-1)}{n^2}\left( {1+\sqrt {1+n^3(n-2)} } \right)$. In Fig.1
it shows
\begin{equation}
\label{eq26}
\frac{(n+1)(n-1)}{n^2}\left( {1+\sqrt {1+n^3(n-2)} } \right)\ge (n-1)^2.
\end{equation}

\begin{figure}[htbp]
\centerline{\includegraphics[width=3in]{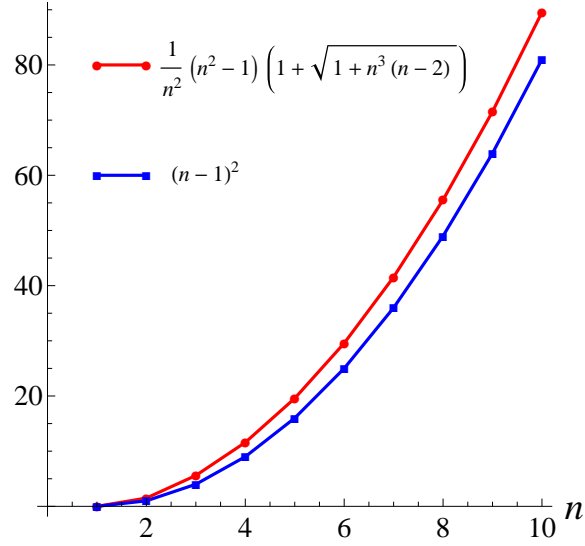}}
\caption{The plot to verify the effective temperature $T_{eff} >0$.}
\label{fig1}
\end{figure}

According to Eq. (\ref{eq23}), $T_{eff} >0$, which fulfills the stability
condition of thermodynamic system. However, the problem of considering the black hole horizon and the cosmological
one as independent each other is when the two horizons coincide , namely
$\kappa _{+} =\kappa _{c}=0$, the temperature on the black hole horizon and the
cosmological horizon are both zero, but the both horizons have nonzero
area, which means that the entropy for the two horizons should not be zero.
The total entropy should be
\[
S=S_+ +S_c =2S_+ =2S_c .
\]
This conclusion is inconsistent with Nernst theorem. In this case the
volume-thermodynamic system becomes area-thermodynamic one. According
to Eq. (\ref{eq23}) the pressure of thermodynamic membrane is zero, but
the temperature of thermodynamic membrane is nonzero. This
can partly solve the problem that extreme de Sitter black holes do not
satisfy the Nernst theorem. When $n=2$, Eqs. (\ref{eq19}) and (\ref{eq20}) return to the
known result.

\section{Phase transition in charged dS black hole spacetime}

The investigation on the phase transition of black holes have
aroused many interest. In particular, the critical behaviors
of some black holes in asymptotically AdS space are similar to ones
of the Van der Waals equation\cite{Banerjee,Kastor,Dolan1,SG,RBM,NA1,Dolan3,DCZ}. For black holes in dS space
there exist more than one horizons and the multiple horizons
correspond to different thermodynamic systems. Thus they are
generally non-equilibrium systems. The research on the phase
transition of this kind of non-equilibrium system is few. Based on
the above section, we analyze the phase transition of higher
dimensional RN-dS black hole. Firstly, we compare the effective
thermodynamic quantities of higher dimensional RN-dS black hole with
the der Waals equation. Secondly, we will discuss the critical
behaviors of thermodynamic quantities of the
RN-dS black hole at constant temperature. At last, the
nature of the phase transition by Ehrenfest's equations will be analyzed.

Comparing with the Van der Waals equation
\begin{equation}
\label{eq27}
\left( {P+\frac{a}{v^2}} \right)(v-b)=kT,
\end{equation}
here, $v=V/N$ is the specific volume of the fluid, $P$ its pressure,
$T$ its temperature, and $k$ is the Boltzmann constant. From
Eq.(\ref{eq27})one can depict the $P-v$ curves for fixed $T$.
Employing the conditions and the equations the critical points
satisfied, one can derive the critical temperature, the critical
pressure and the critical volume. To compare with the Van der Waals
equation,we set $P_{eff} \to P$ and discuss the phase
transition and critical phenomena when $Q$ is invariant.

Substituting Eq.(\ref{eq19}) into Eq.(\ref{eq20}), we have
\begin{equation}
\label{eq28}
P_{eff} =T_{eff} \frac{B_4 }{2r_c B_2 }+\frac{B_2 B_3 -B_1 B_4 }{8\pi r_c^2
x(1+x)B_2 },
\end{equation}
in which $x$ is a dimensionless
parameter. We take the specific volume of the higher dimensional RN-dS spacetime as
\begin{equation}
\label{eq29}
v=r_c (1-x).
\end{equation}
According to
\begin{equation}
\label{eq30}
\left( {\frac{\partial P_{eff} }{\partial v}} \right)_{T_{eff} } =0,
\quad
\left( {\frac{\partial ^2P_{eff} }{\partial v^2}} \right)_{T_{eff} } =0,
\end{equation}
we can calculate the position of the critical points.

Here we take $Q=1,~3,~10$ respectively and calculate the various quantities at the critical
points in different spacetime dimensions ($n=3,~4,~5)$. The results are shown in Table I.

\begin{table}[!htbp]
\begin{center}
\begin{tabular}{|c|c|c|c|c|c|c|c|c|c|}
\hline
\raisebox{-1.50ex}[0cm][0cm]{}&
\multicolumn{3}{|c|}{$n=3$} &
\multicolumn{3}{|c|}{$n=4$} &
\multicolumn{3}{|c|}{$n=5$}  \\
\cline{2-10}
&
$Q=1$&
$Q=3$&
$Q=10$&
$Q=1$&
$Q=3$&
$Q=10$&
$Q=1$&
$Q=3$&
$Q=10$ \\
\hline
$x^c$&
0.769263&
0.769263&
0.769263&
0.798722&
0.798722&
0.798722&
0.821785&
0.821785&
0.821785 \\
\hline
$r_c^c $&
1.61337&
2.79444&
5.10192&
1.31405&
1.89519&
2.83103&
1.21688&
1.60151&
2.16396 \\
\hline
$T_{eff}^c $&
0.030460&
0.017586&
0.009232&
0.047178&
0.032711&
0.021898&
0.057924&
0.044013&
0.032573 \\
\hline
$P_{eff}^c $&
0.008375&
0.002791&
0.000837&
0.022521&
0.010827&
0.004852&
0.038737&
0.022365&
0.012249 \\
\hline $v^c$& 0.372263& 0.644779& 1.1772& 0.264489& 0.381459&
0.569823& 0.216866& 0.285412&
0.385648 \\
\hline
$V^c$&
21.7266&
195.54&
2172.66&
13.9192&
86.8592&
646.071&
11.6117&
60.336&
367.193 \\
\hline
$M^c$&
1.26609&
3.79828&
12.6609&
1.38401&
4.15202&
13.8401&
1.51006&
4.5319&
15.1006 \\
\hline
$S^c$&
30.1578&
156.705&
953.675&
27.6023&
119.428&
594.674&
28.436&
112.272&
505.672 \\
\hline
$G^c$&
0.52945&
1.58835&
5.2945&
0.395263&
1.18579&
3.95263&
0.312726&
0.938178&
3.12726 \\
\hline
\end{tabular}
\label{tab1}
\end{center}
\caption{Numerical solutions for $x^c$, $r_c^c$,  $T_{eff}^c$, $P_{eff}^c$, $v^c$, $V^c$, $M^c$, $S^c$ and $G^c$ for given values of $Q=1, 3, 10$ and spacetime
dimension $n=3,4,5$ respectively.}
\end{table}

According to the above calculations, the position $x^c$ of the
critical point of the higher dimensional RN-dS black holes is
independent of the electric charge. However, it is relevant to the
dimension of spacetime and increases with the dimension. The
position of the critical cosmological horizon $r_c^c $ increases
with the electric charge $Q$ for fixed spacetime dimension, and
decrease with the increase of the spacetime dimension for fixed
electric charge. Both the critical effective temperature $T_{eff}^c$
and the critical effective pressure $P_{eff}^c $ decrease with the increase of the electric charge $Q$ for fixed
spacetime dimension, and increase with the spacetime dimension for
fixed electric charge. The critical specific volume $v^c$ increase
with the increase of the electric charge $Q$ for fixed spacetime
dimension, and decrease with the spacetime dimension for fixed
electric charge.

In order to describe the relation of $P_{eff} $ and $v$ in the vicinity of critical temperature, we plot the curves of $P_{eff} $-$v$ at different temperatures. For the sake of brevity, we only depict the curves for the $n=5$
case. In the cases with $n=3,~4$ the  behaviors are similar.

\begin{figure}[!htbp]
\center{\subfigure[~$ Q=1 $] {
\includegraphics[angle=0,width=5cm,keepaspectratio]{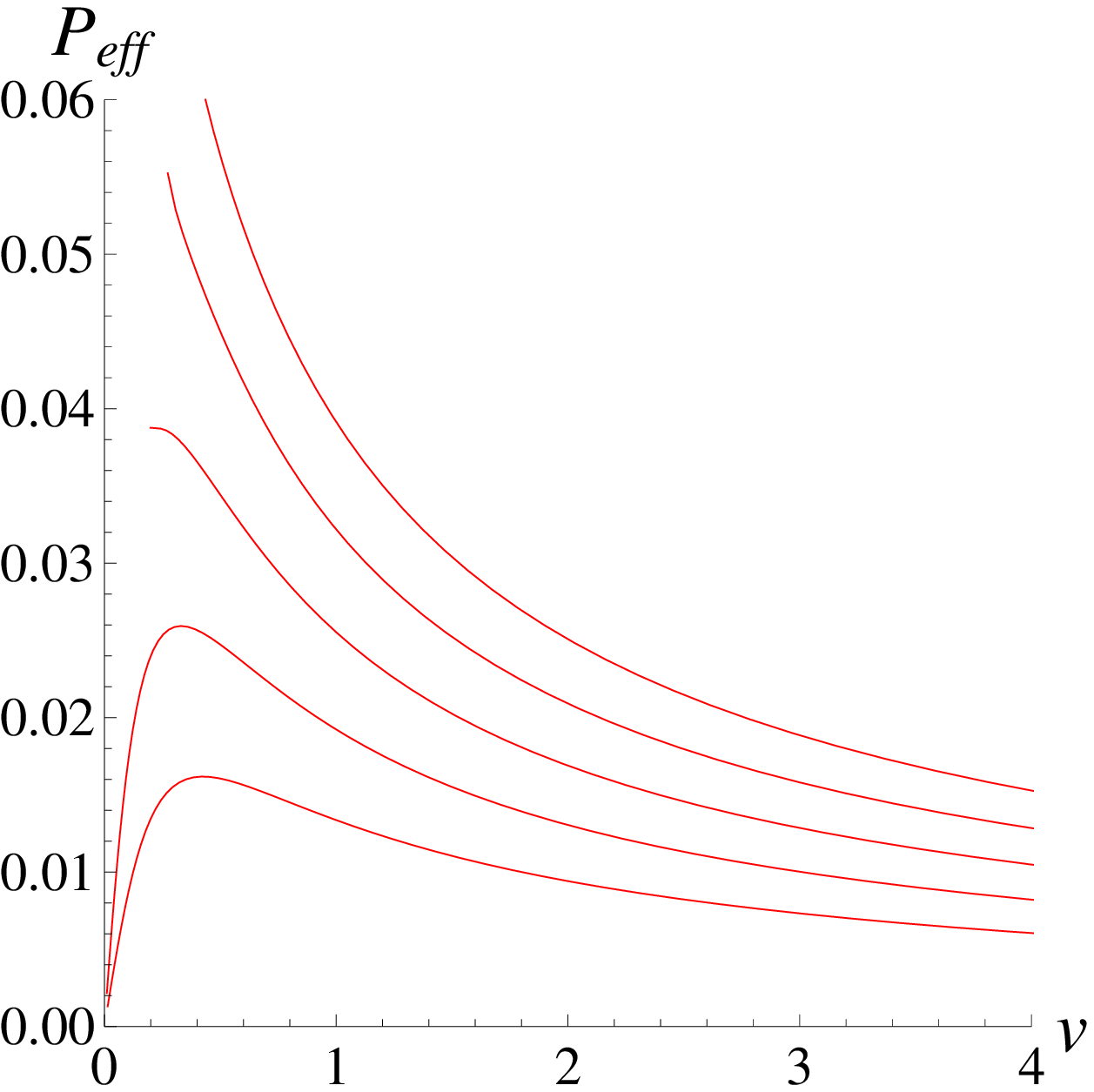}}
\subfigure[~$ Q=3 $] {
\includegraphics[angle=0,width=5cm,keepaspectratio]{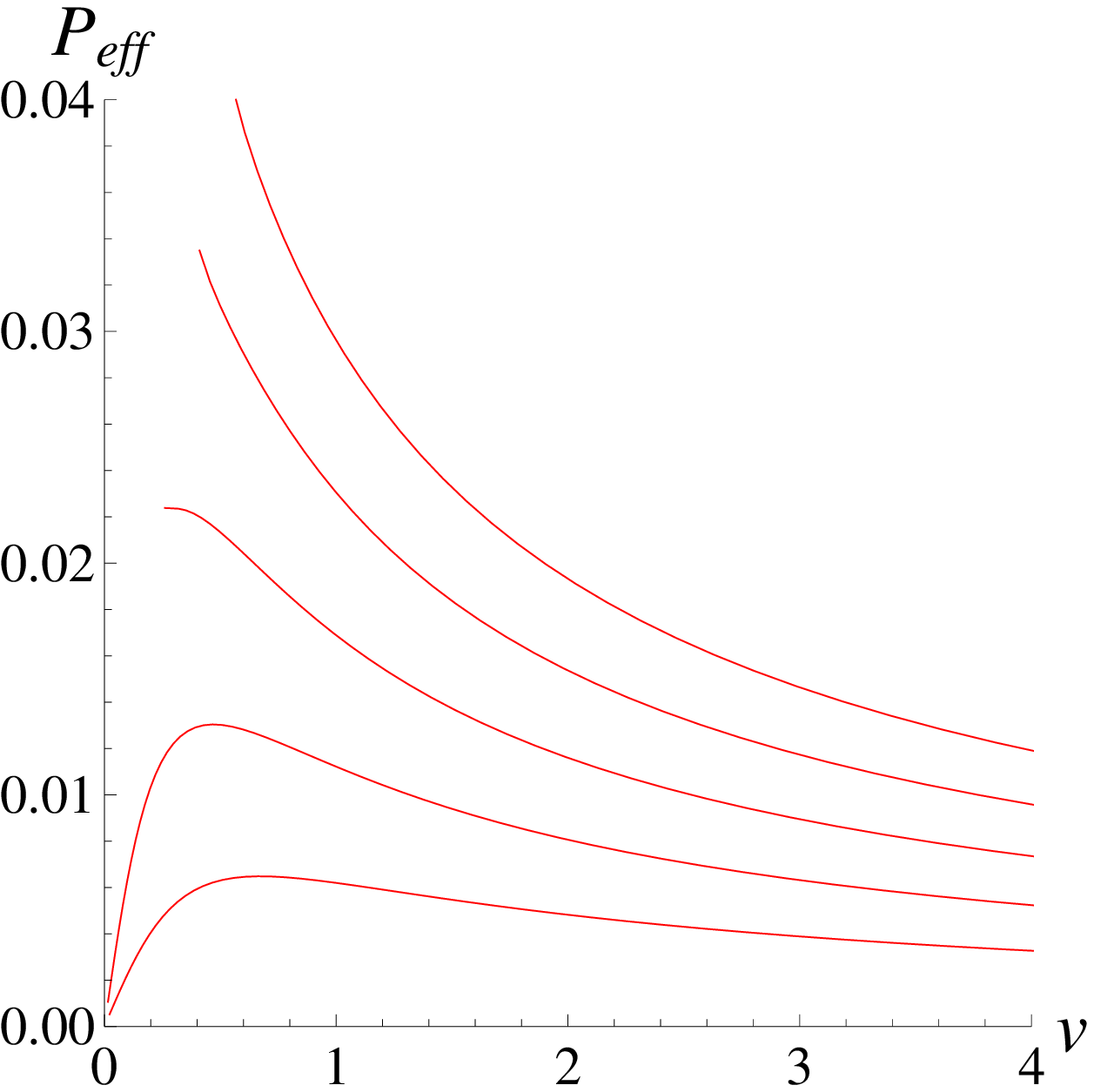}}
\subfigure[~$ Q=10 $] {
\includegraphics[angle=0,width=5cm,keepaspectratio]{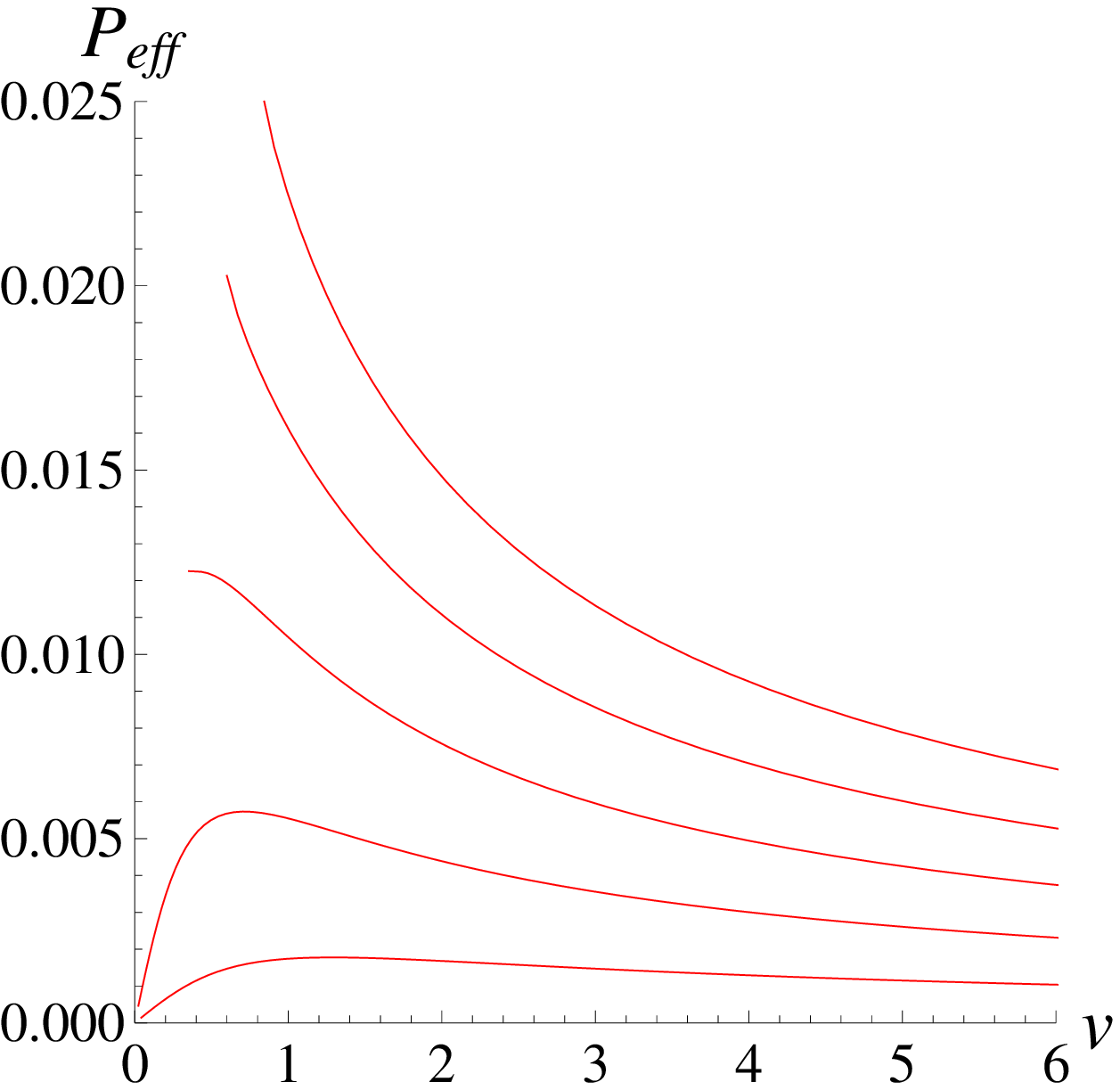}}
\caption[]{\it The $P_{eff}-v$ curves for $Q=1, 3, 10$ respectively.
From top to the bottom the curves correspond to the effective
temperature $T_{eff}^c +0.02$, $T_{eff}^c +0.01$, $T_{eff}^c $,
$T_{eff}^c -0.01$ and $T_{eff}^c -0.02$.}} \label{pv}
\end{figure}

From Figs.2, when the effective temperature $T_{eff} >T_{eff}^c $,
the stable condition $\left( {\frac{\partial P}{\partial v}}
\right)_{T_{eff} } <0$ can be satisfied. In the case of  $T_{eff}
<T_{eff}^c $, only when the value of $v$ is large, the stable
condition $\left( {\frac{\partial P}{\partial v}} \right)_{T_{eff} }
<0$ can be satisfied. When the system lies at a small $v$, $\left(
{\frac{\partial P}{\partial v}} \right)_{T_{eff} }
>0$, thus the system is unstable. So phase transition may occur only at $T_{eff} =T_{eff}^c $.

\section{Analytical check of the classical Ehrenfest equations in the extended
phase space}

According to Ehrenfest's classification, when the chemical potential and its first derivative are continuous, whereas
the second derivative of chemical potential is discontinuous, this kind of
phase transition is called the second-order phase transition.
For Van der Waals system there is no latent heat and the liquid-gas structure do not change suddenly at the critical
point. Therefor this kind of phase transition belongs to the second-order phase transition and continuous phase
transition . Below we will discuss the behaviors of higher dimensional RN-dS system near the
phase transition point.

We can calculate the specific heat of RN-dS system at constant pressure $C_P $, the
volume expansivity $\beta $, and the isothermal compressibility $\kappa $
\[
C_P =T_{eff} \left( {\frac{\partial S}{\partial T_{eff} }} \right)_{P_{eff}
} =-T_{eff} \frac{\partial ^2\mbox{G}}{\partial T_{eff}^2 }
\]
\begin{equation}
\label{eq33}
=\frac{Vol(S^n)r_c^{n-1} }{4}nT_{eff}
\left( {\frac{r_c (1+x^{n-1})\left( {\frac{\partial P_{eff} }{\partial r_c
}} \right)_x -(1+x^n)\left( {\frac{\partial P_{eff} }{\partial x}}
\right)_{r_c } }{\left( {\frac{\partial T_{eff} }{\partial x}} \right)_{r_c
} \left( {\frac{\partial P_{eff} }{\partial r_c }} \right)_x -\left(
{\frac{\partial T_{eff} }{\partial r_c }} \right)_x \left( {\frac{\partial
P_{eff} }{\partial x}} \right)_{r_c } }} \right),
\end{equation}
\[
\beta =\frac{1}{v}\left( {\frac{\partial v}{\partial T_{eff} }}
\right)_{P_{eff} } =\frac{1}{v}\frac{\partial ^2\mu }{\partial T_{eff}
\partial P_{eff} }
\]
\begin{equation}
\label{eq34}
=-\frac{1}{v}\left( {\frac{r_c \left( {\frac{\partial P_{eff} }{\partial r_c
}} \right)_x +(1-x)\left( {\frac{\partial P_{eff} }{\partial x}}
\right)_{r_c } }{\left( {\frac{\partial T_{eff} }{\partial x}} \right)_{r_c
} \left( {\frac{\partial P_{eff} }{\partial r_c }} \right)_x -\left(
{\frac{\partial T_{eff} }{\partial r_c }} \right)_x \left( {\frac{\partial
P_{eff} }{\partial x}} \right)_{r_c } }} \right),
\end{equation}
\[
\kappa =-\frac{1}{v}\left( {\frac{\partial v}{\partial P_{eff} }}
\right)_{T_{eff} } =-\frac{1}{v}\frac{\partial ^2\mu }{\partial P_{eff}^2 }
\]
\begin{equation}
\label{eq35}
=-\frac{1}{v}\left( {\frac{r_c \left( {\frac{\partial T_{eff} }{\partial r_c
}} \right)_x +(1-x)\left( {\frac{\partial T_{eff} }{\partial x}}
\right)_{r_c } }{\left( {\frac{\partial P_{eff} }{\partial r_c }} \right)_x
\left( {\frac{\partial T_{eff} }{\partial x}} \right)_{r_c } -\left(
{\frac{\partial P_{eff} }{\partial x}} \right)_{r_c } \left( {\frac{\partial
T_{eff} }{\partial r_c }} \right)_x }} \right),
\end{equation}
where $S=\frac{Vol(S^n)}{4}r_c^n (1+x^n)$.

We also depict the curves of $C_P -x$, $\beta -x$ and $\kappa -x$ in the FIG 3, FIG 4 and FIG 5 respectively.
From these curves, we find that the specific heat of higher dimensional RN-dS system at constant pressure $C_P $,
the expansion coefficient $\beta $, and the compressibility $\kappa $
exist infinite peak. While the curves of $S -x$ and $G -x$ in FIG 6 and FIG 7 show that the Gibbs function
$G$ and the entropy $S$ are both continuous at the critical point. According
to Ehrenfest, the phase transition of the higher dimensional RN-dS black
hole should be the second-order one, which is similar to the four
dimensional case.

\begin{figure}[!htbp]
\center{\subfigure[~$ Q=1 $] {
\includegraphics[angle=0,width=5cm,keepaspectratio]{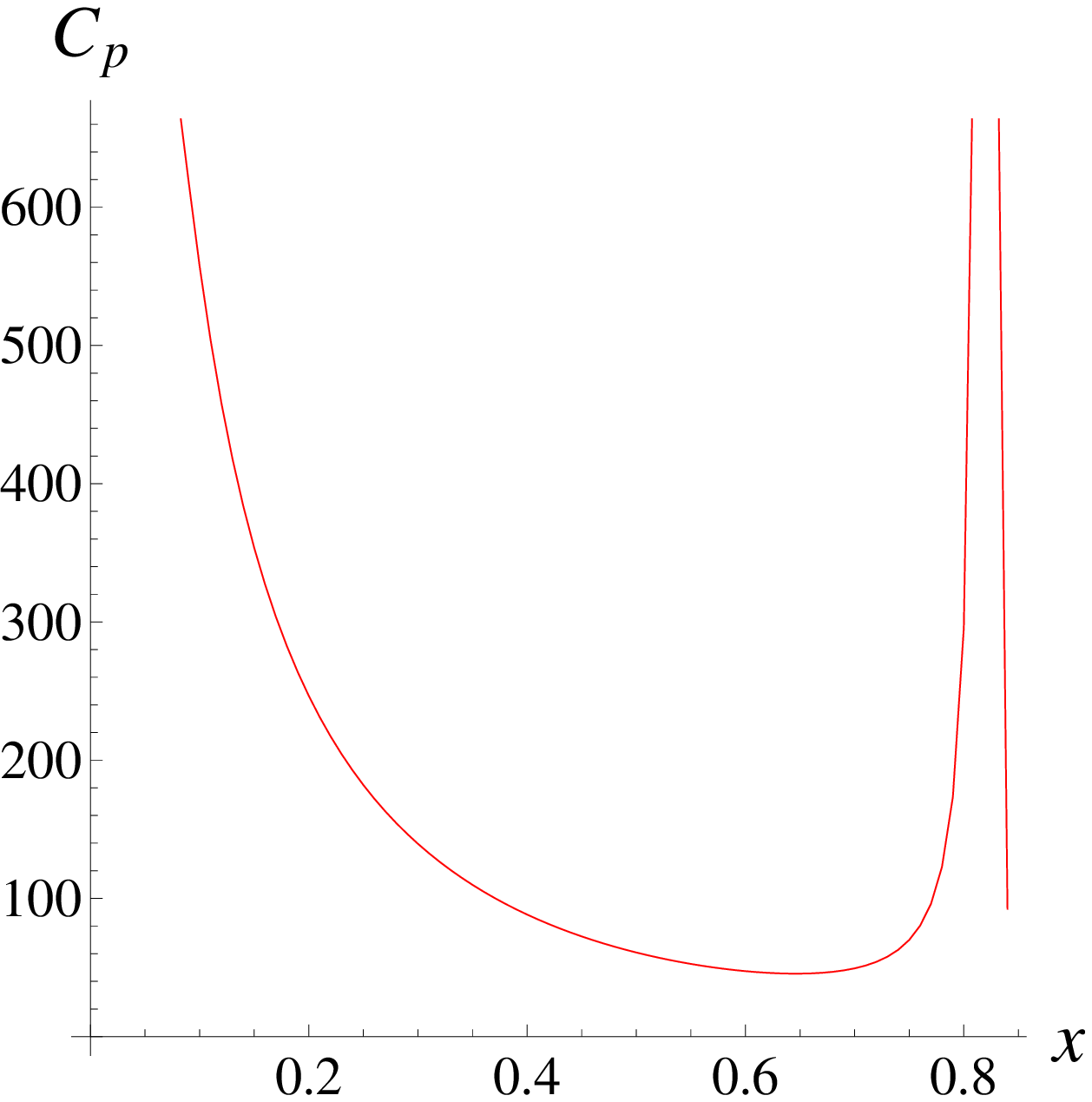}}
\subfigure[~$ Q=3 $] {
\includegraphics[angle=0,width=5cm,keepaspectratio]{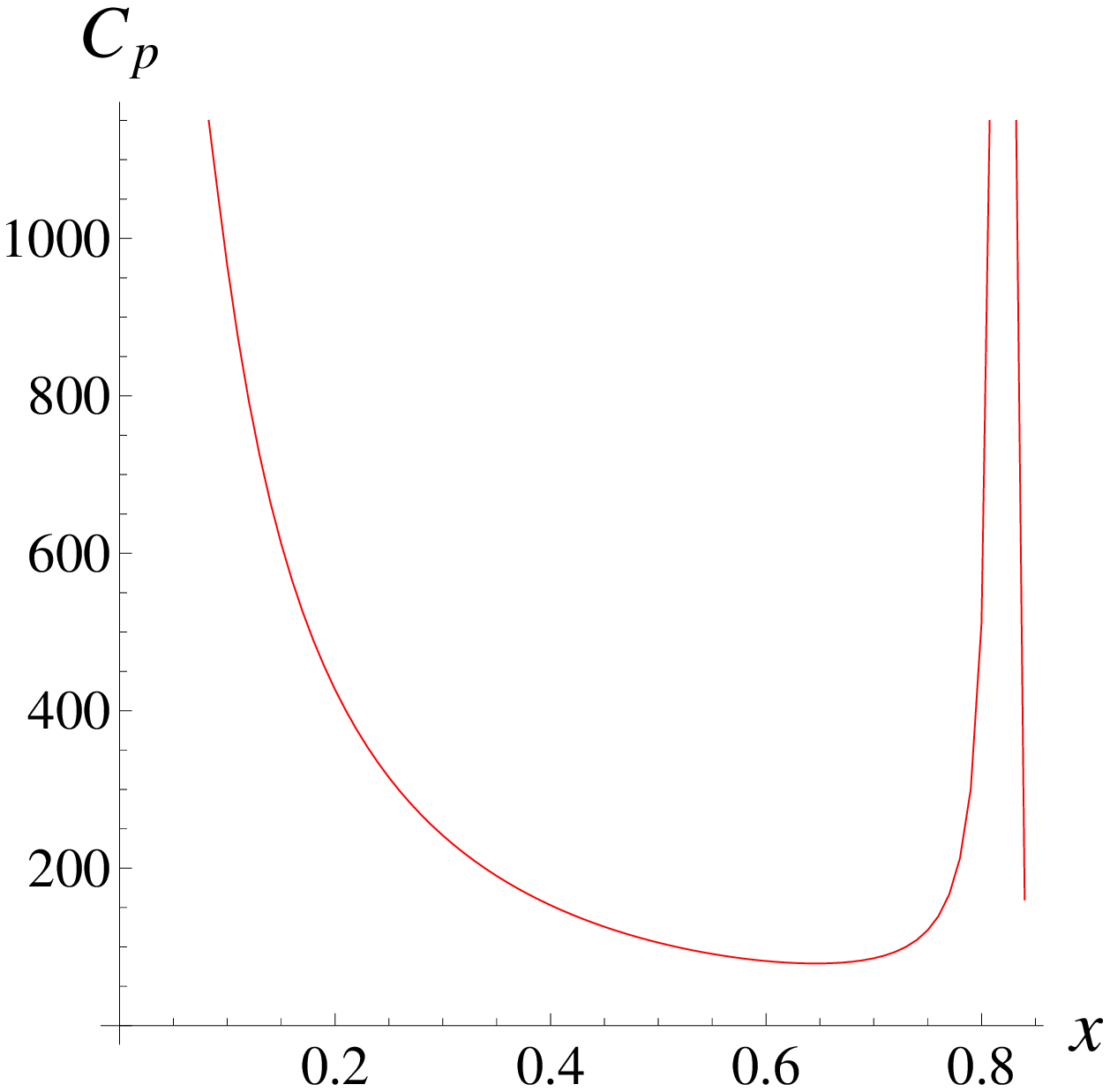}}
\subfigure[~$ Q=10 $] {
\includegraphics[angle=0,width=5cm,keepaspectratio]{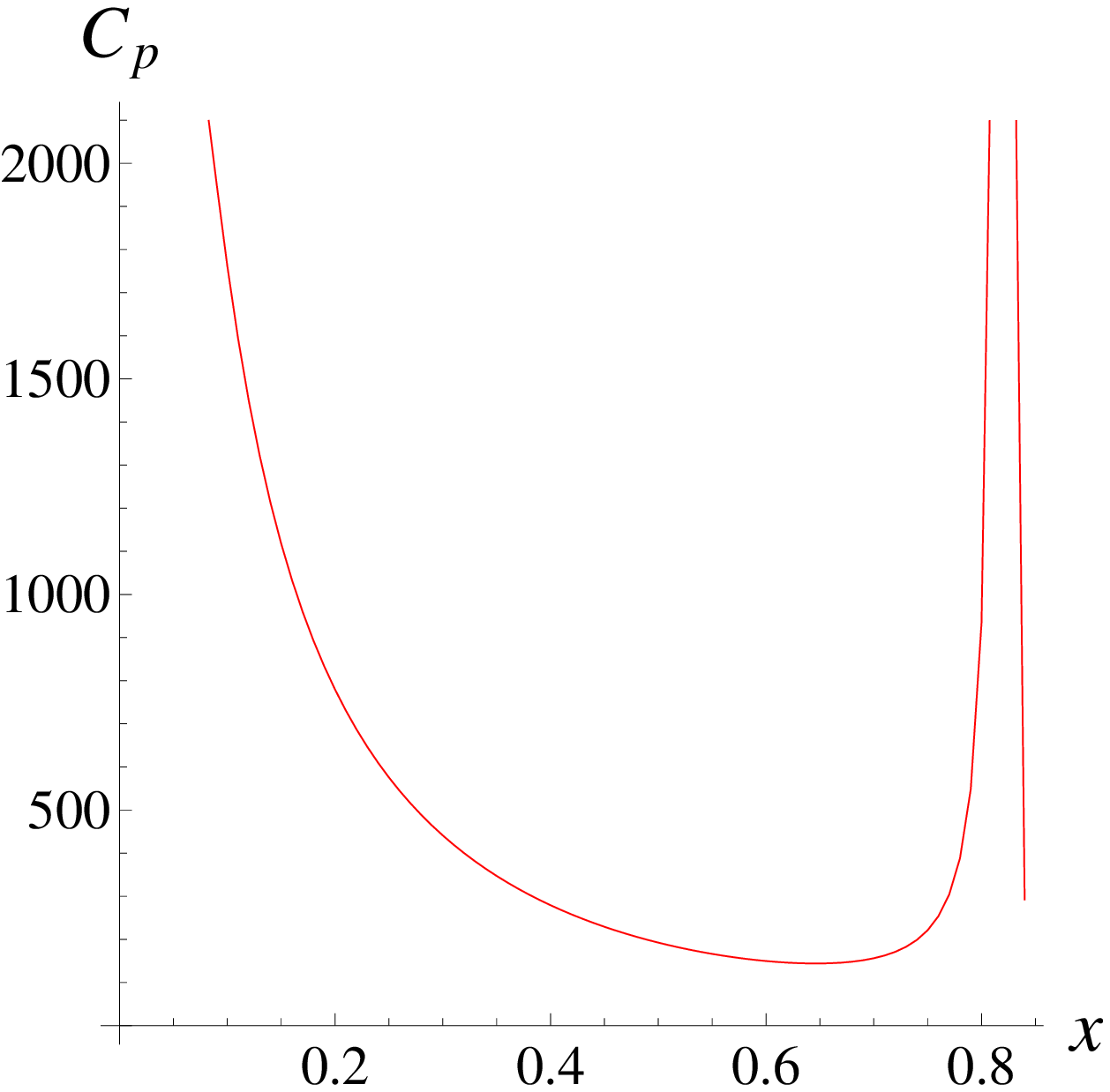}}
\caption[]{\it $C_P-x$ curves for RN-dS black hole with $n=5$ corresponding to
the critical effective pressure  $P_{eff}^c =0.038737$,
$P_{eff}^c =0.022365$ and $P_{eff}^c =0.012249$ respectively.}}
\label{Cx}
\end{figure}

\begin{figure}[!htbp]
\center{\subfigure[~$ Q=1 $] {
\includegraphics[angle=0,width=5cm,keepaspectratio]{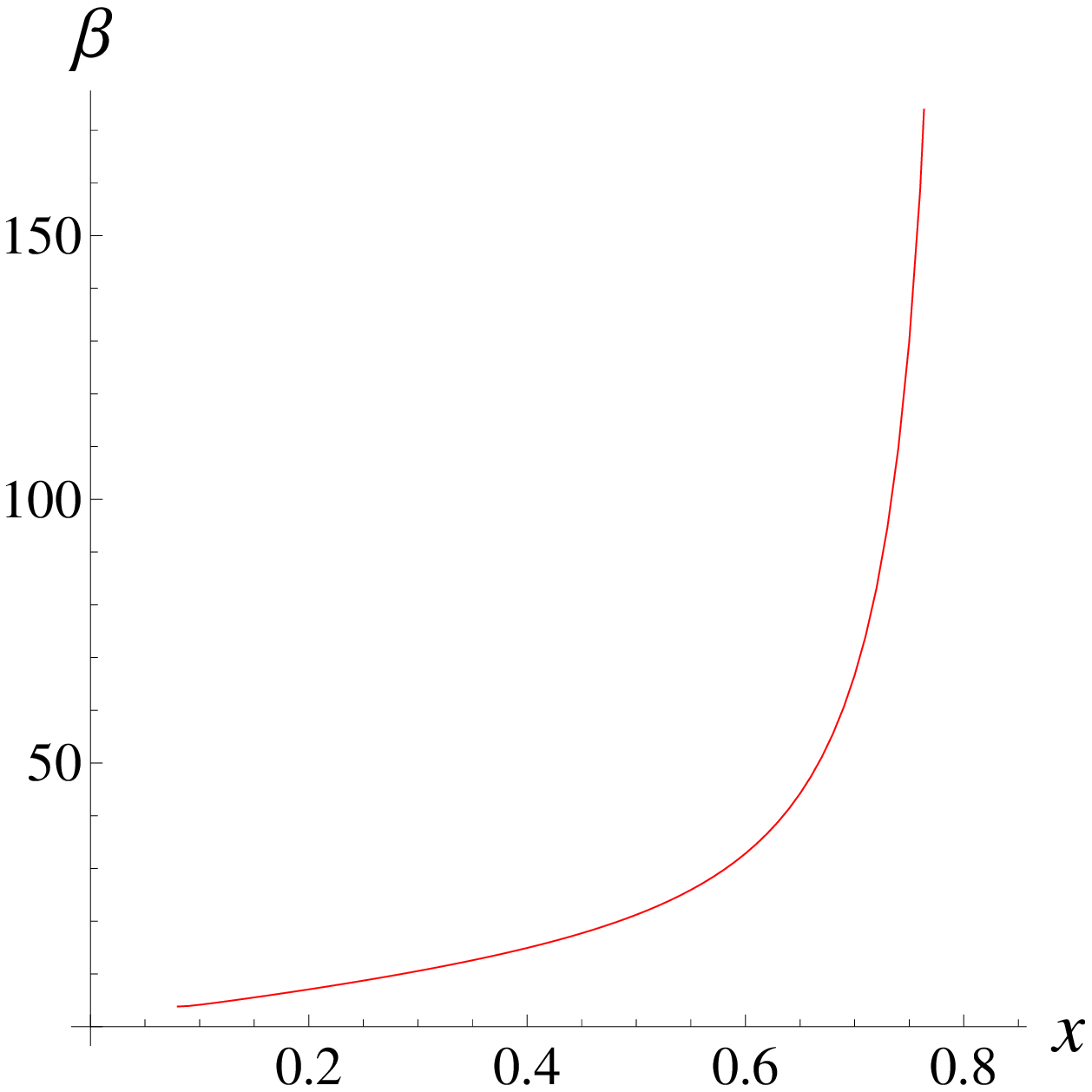}}
\subfigure[~$ Q=3 $] {
\includegraphics[angle=0,width=5cm,keepaspectratio]{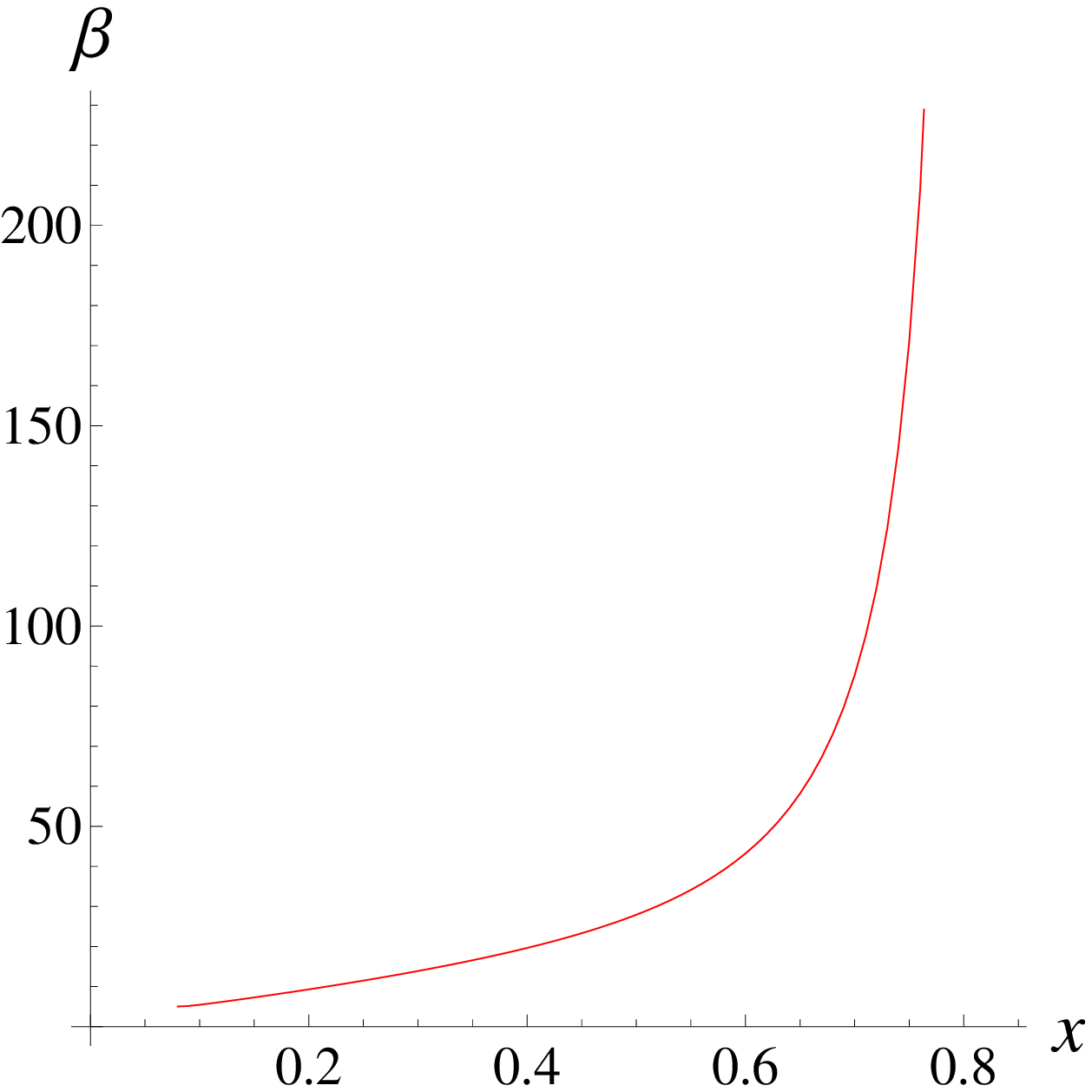}}
\subfigure[~$ Q=10 $] {
\includegraphics[angle=0,width=5cm,keepaspectratio]{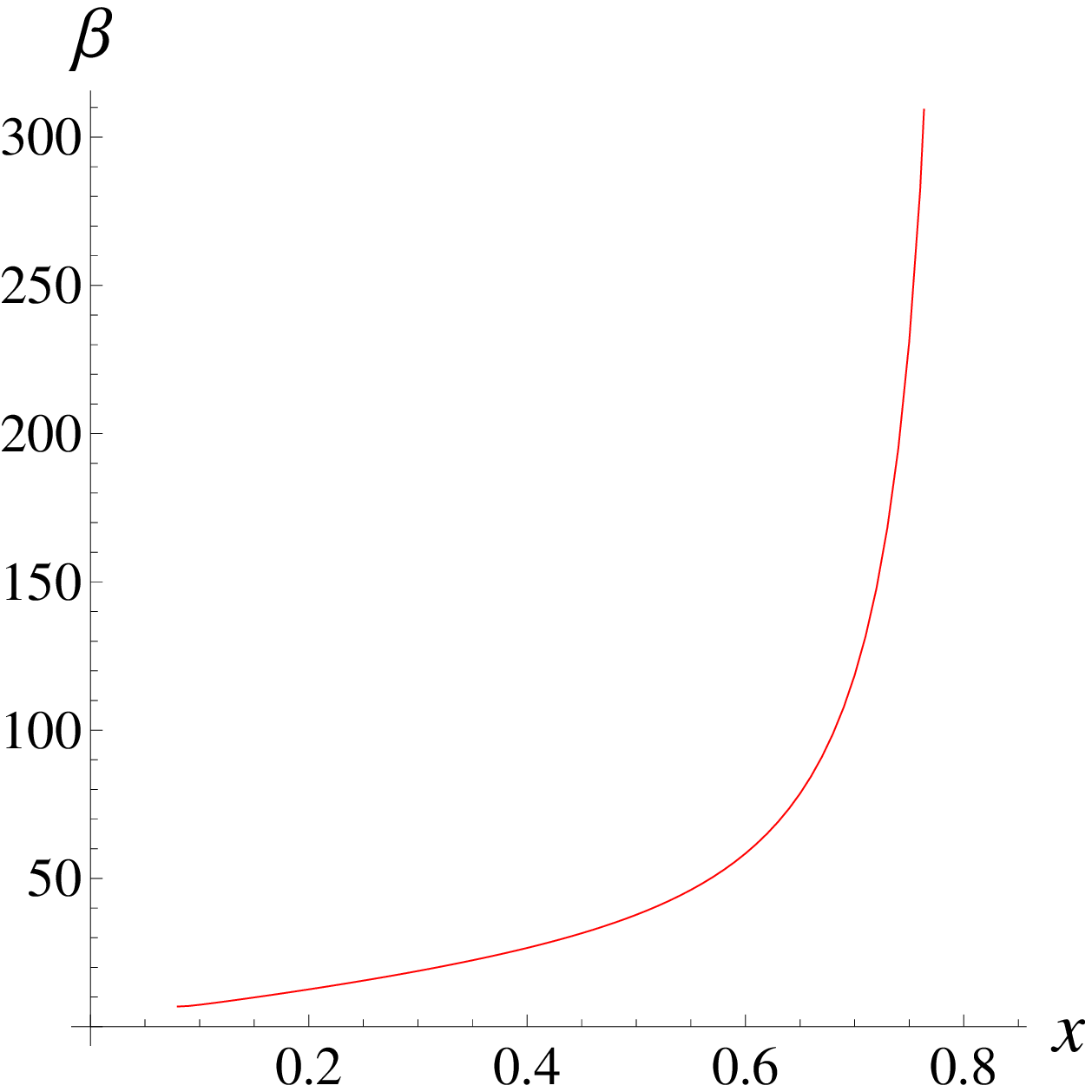}}
\caption[]{\it $\beta-x$ curves for RN-dS black hole with $n=5$ corresponding
to the critical effective pressure  $P_{eff}^c =0.038737$,
$P_{eff}^c =0.022365$ and $P_{eff}^c =0.012249$
respectively.}}
\label{bx}
\end{figure}

\begin{figure}[!htbp]
\center{\subfigure[~$ Q=1 $] {
\includegraphics[angle=0,width=5cm,keepaspectratio]{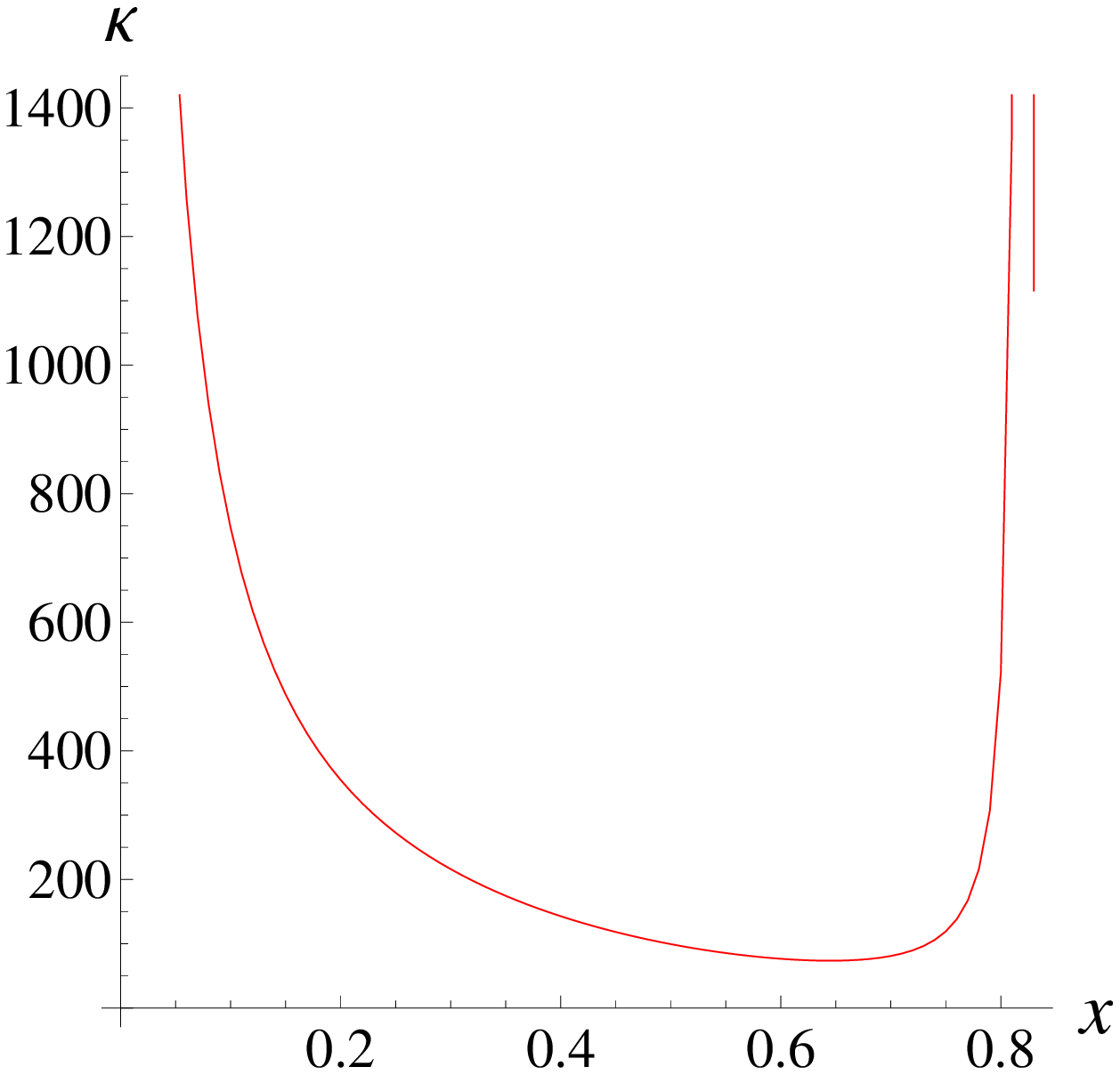}}
\subfigure[~$ Q=3 $] {
\includegraphics[angle=0,width=5cm,keepaspectratio]{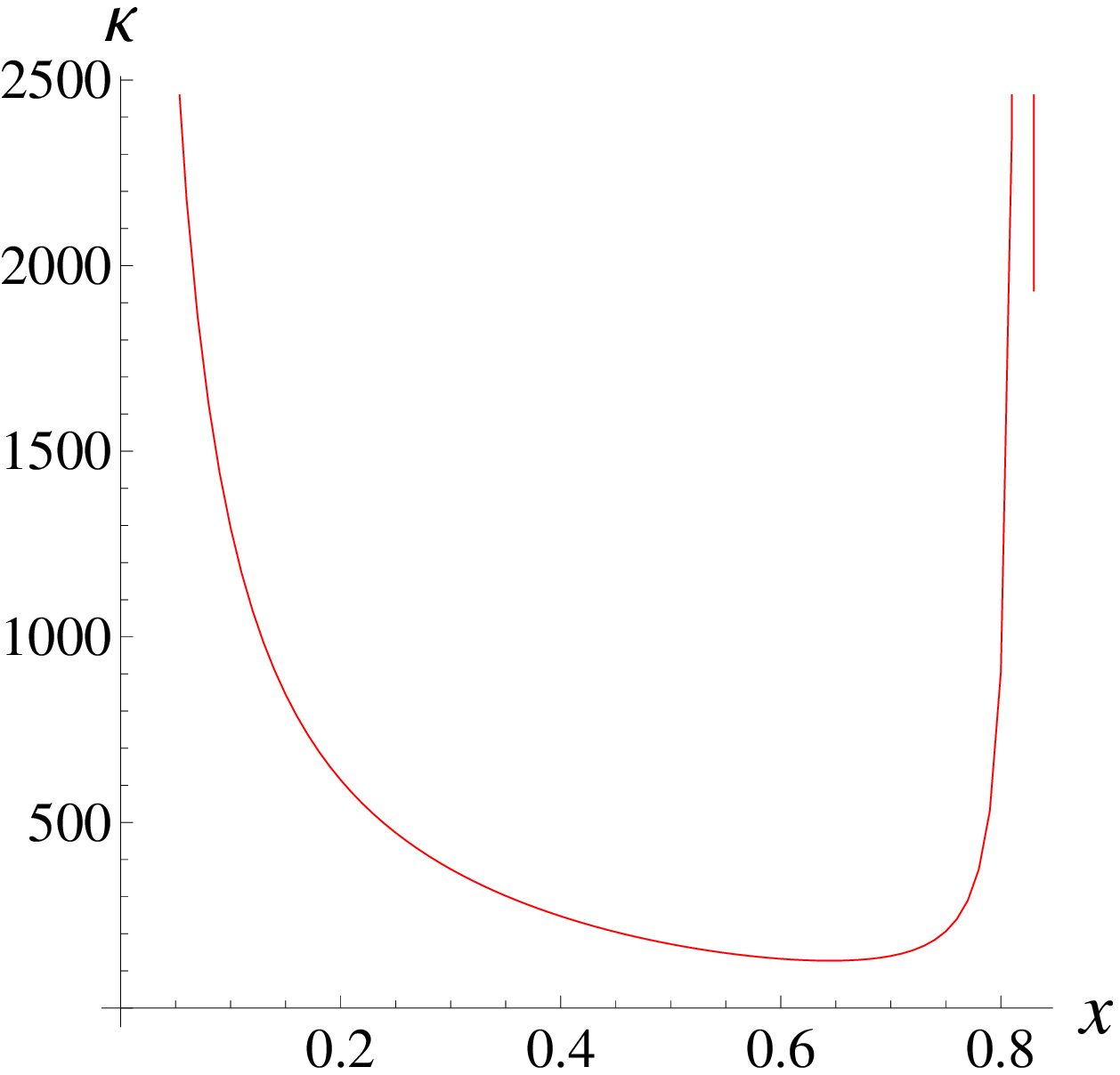}}
\subfigure[~$ Q=10 $] {
\includegraphics[angle=0,width=5cm,keepaspectratio]{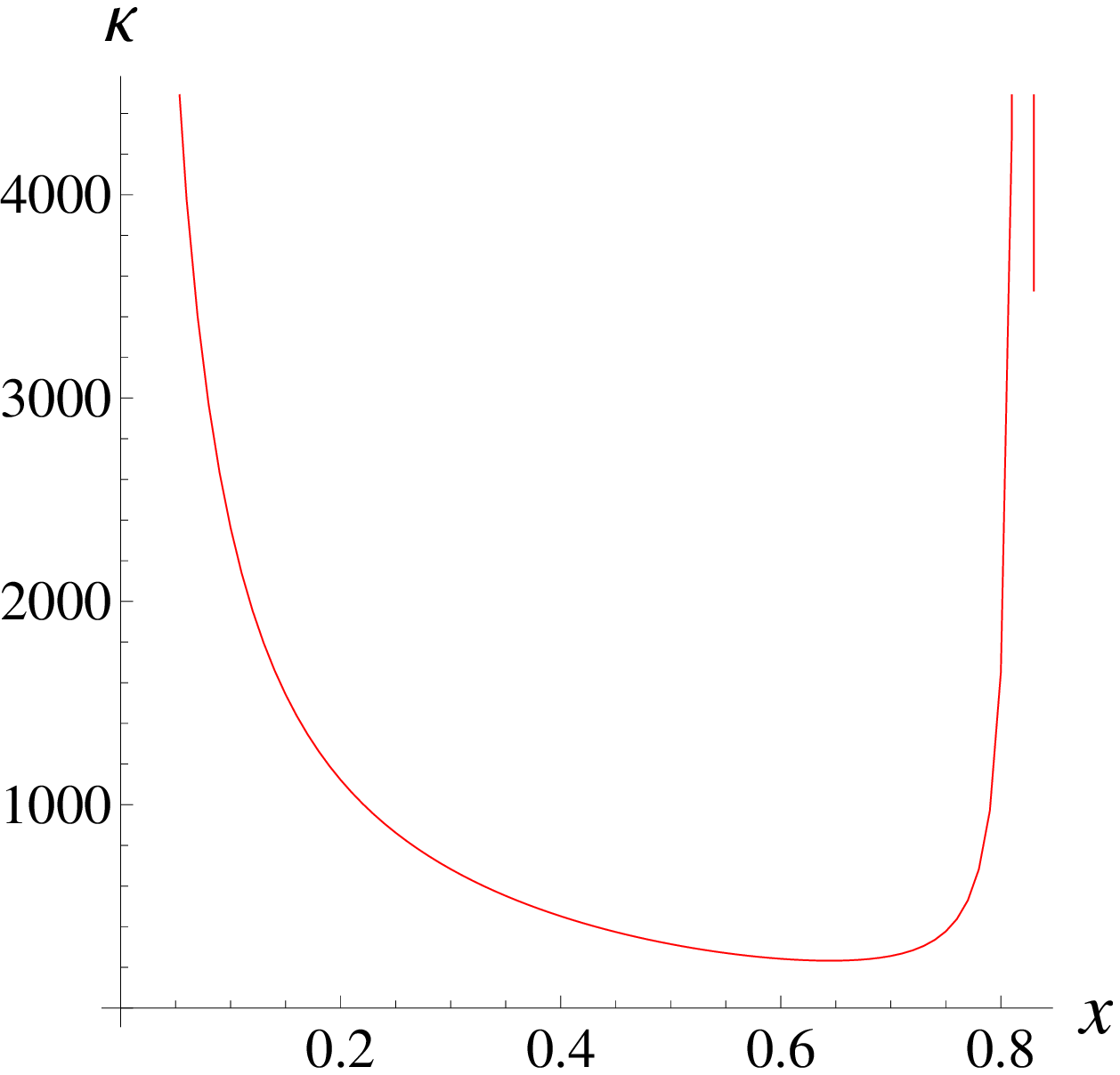}}
\caption[]{\it $\kappa-x$ curves for RN-dS black hole with $n=5$ corresponding
to the critical effective temperature  $T_{eff}^c =0.057924$,
$T_{eff}^c =0.044013$ and $T_{eff}^c =0.032573$ respectively.}}
\label{kx}
\end{figure}

\begin{figure}[!htbp]
\center{\subfigure[~$ Q=1 $] {
\includegraphics[angle=0,width=5cm,keepaspectratio]{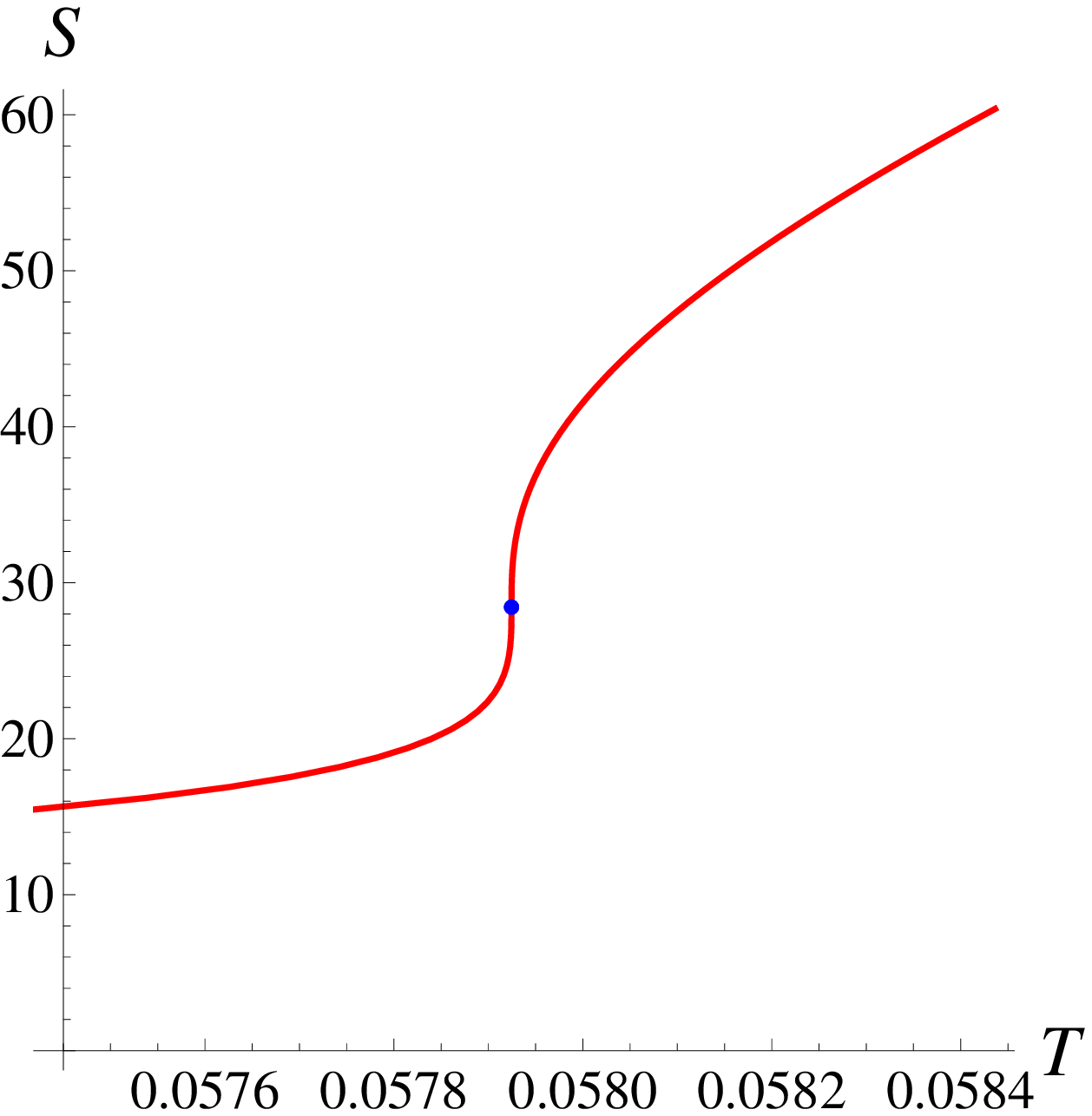}}
\subfigure[~$ Q=3 $] {
\includegraphics[angle=0,width=5cm,keepaspectratio]{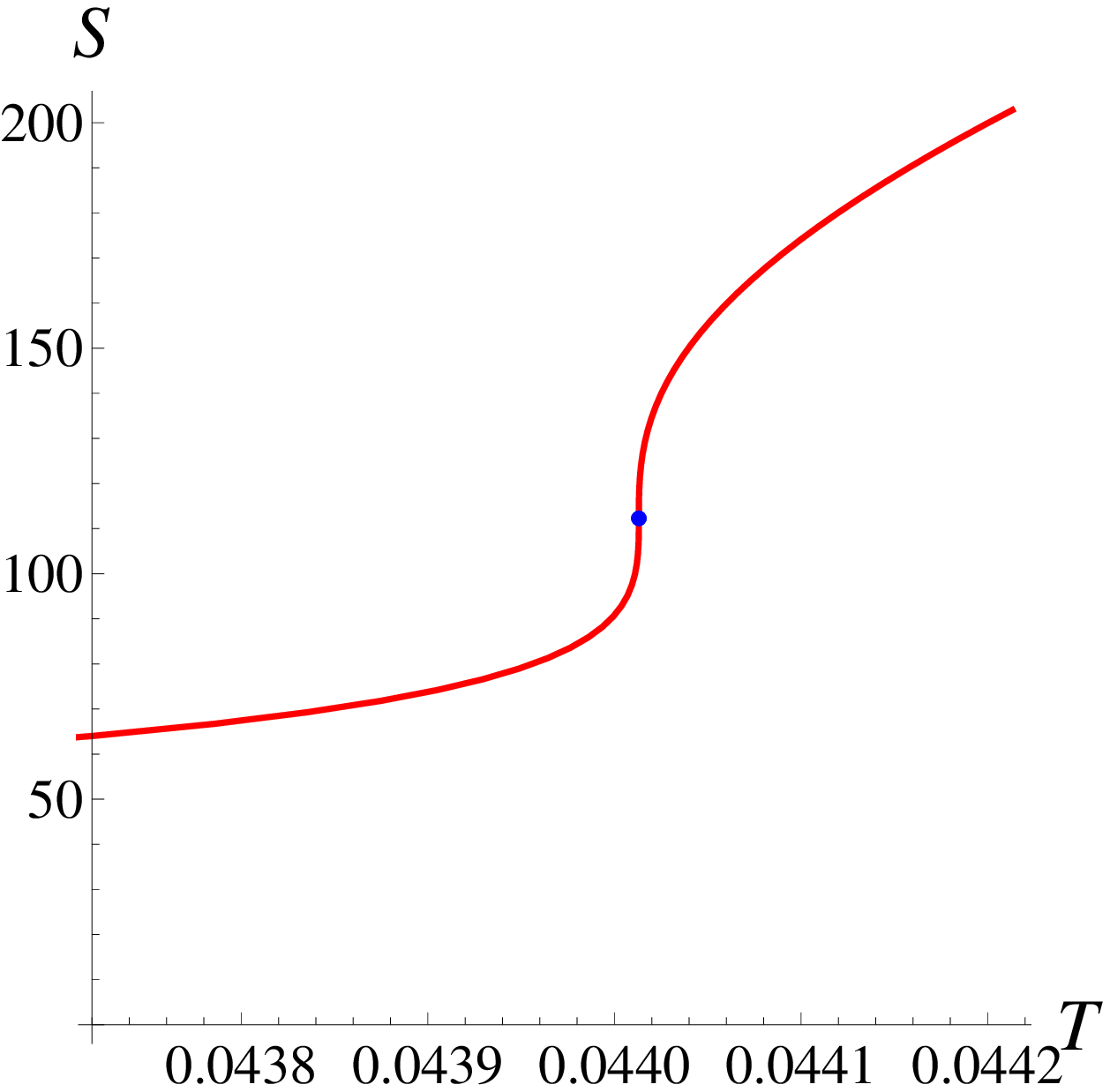}}
\subfigure[~$ Q=10 $] {
\includegraphics[angle=0,width=5cm,keepaspectratio]{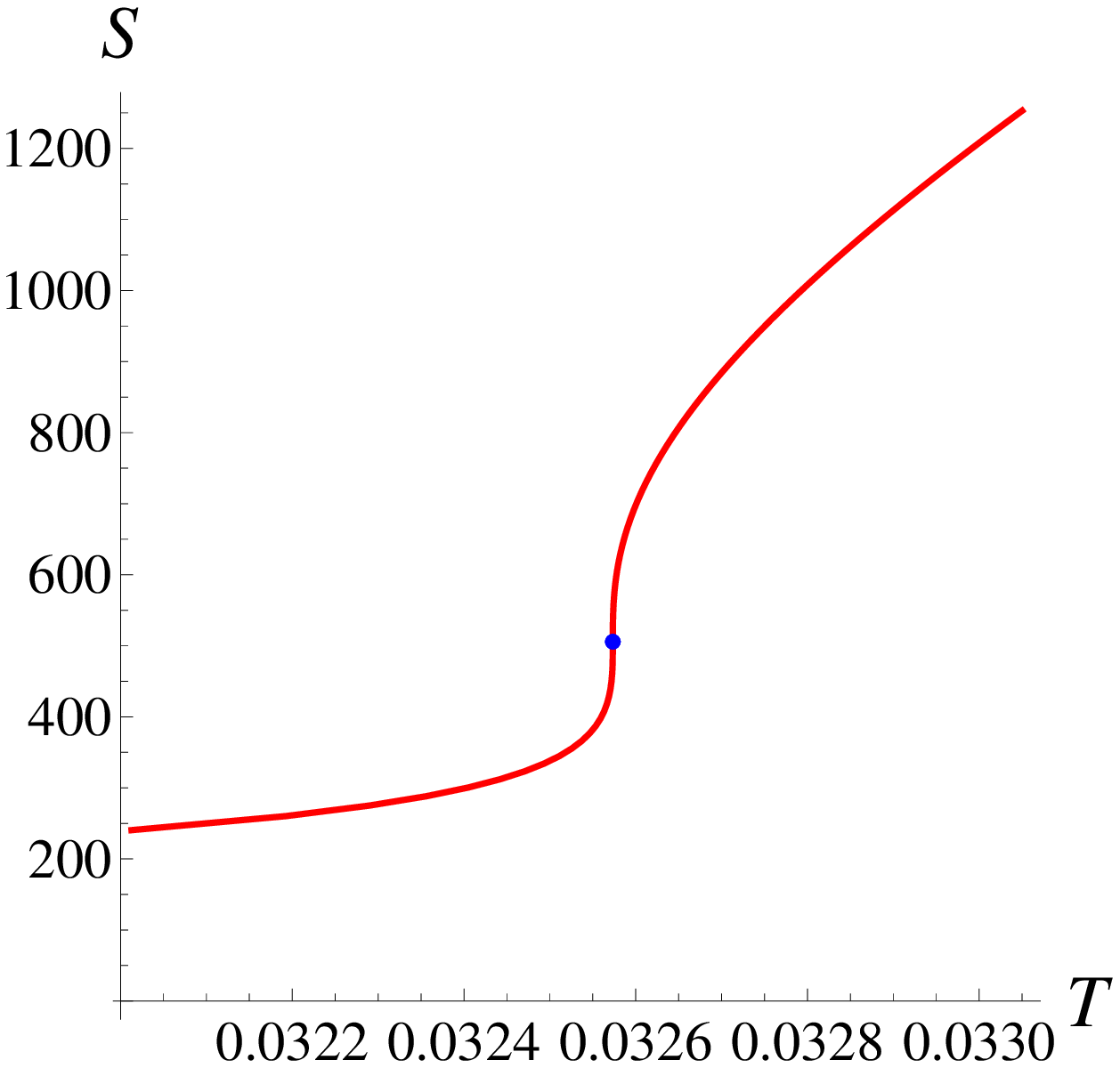}}
\caption[]{\it $S-T$ curves for RN-dS black hole with $n=5$ corresponding to
the critical effective pressure  $P_{eff}^c =0.038737$,
$P_{eff}^c =0.022365$ and $P_{eff}^c =0.012249$ respectively.}}
\label{ST}
\end{figure}

\begin{figure}[!htbp]
\center{\subfigure[~$ Q=1 $] {
\includegraphics[angle=0,width=5cm,keepaspectratio]{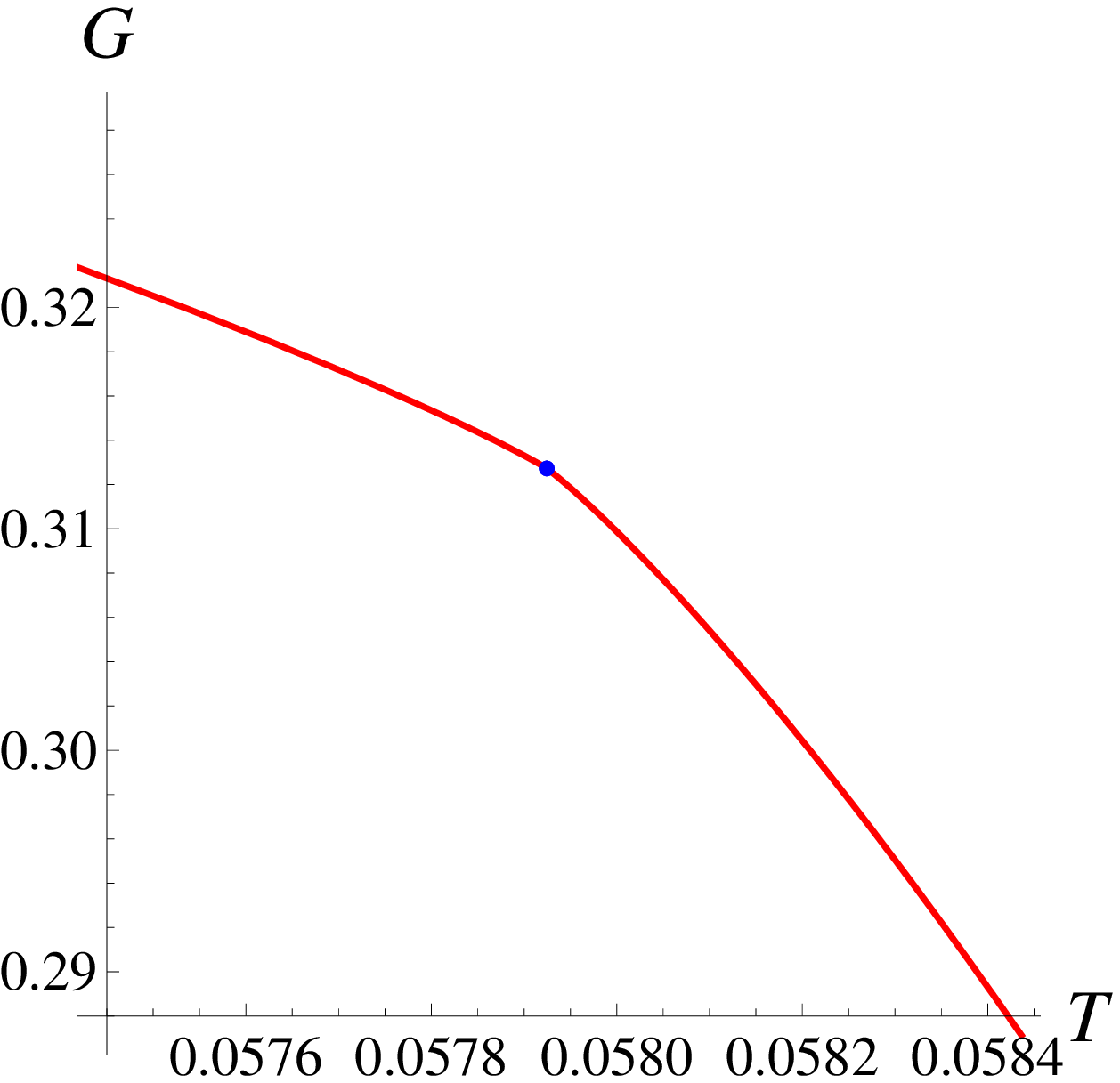}}
\subfigure[~$ Q=3 $] {
\includegraphics[angle=0,width=5cm,keepaspectratio]{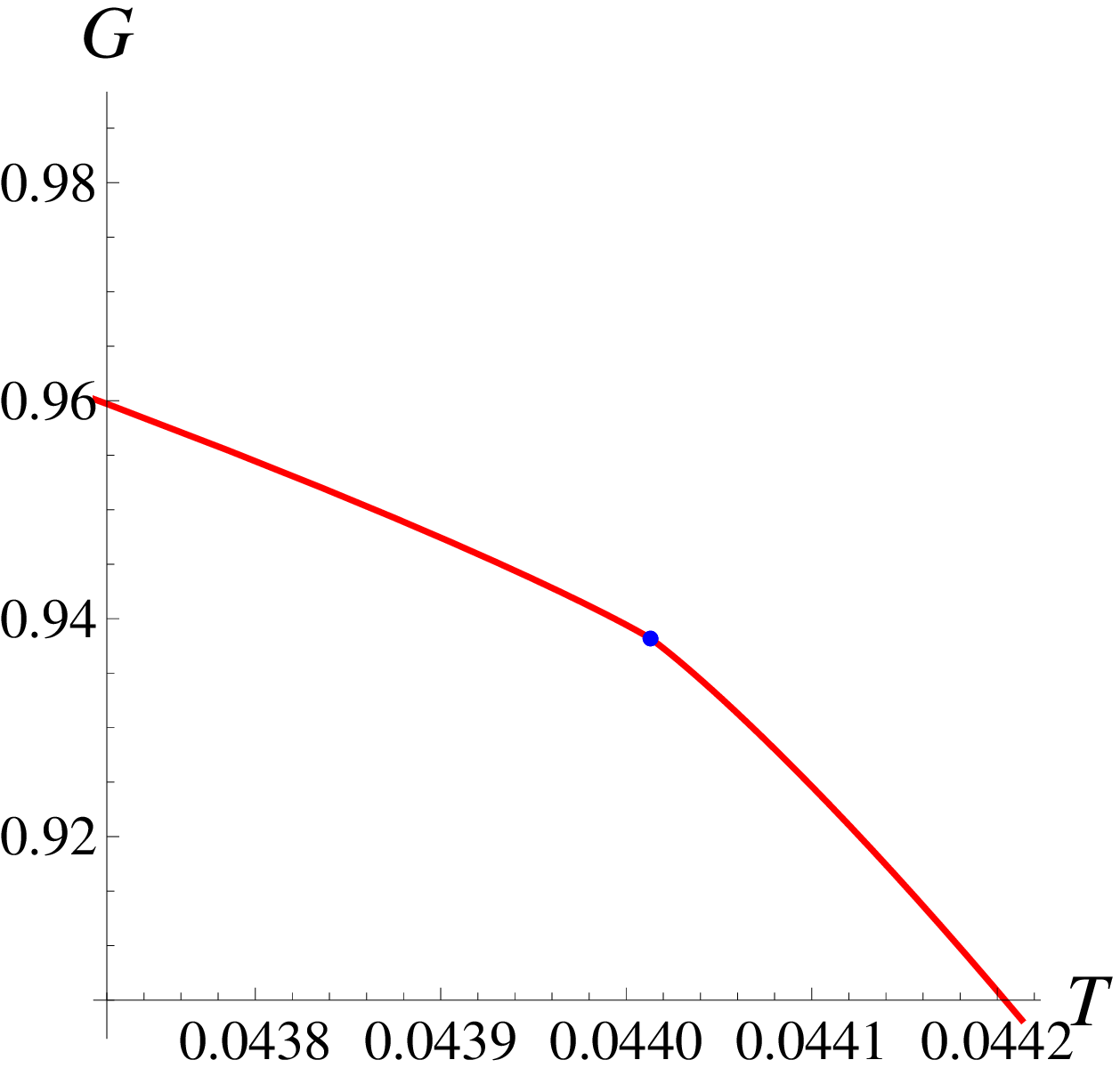}}
\subfigure[~$ Q=10 $] {
\includegraphics[angle=0,width=5cm,keepaspectratio]{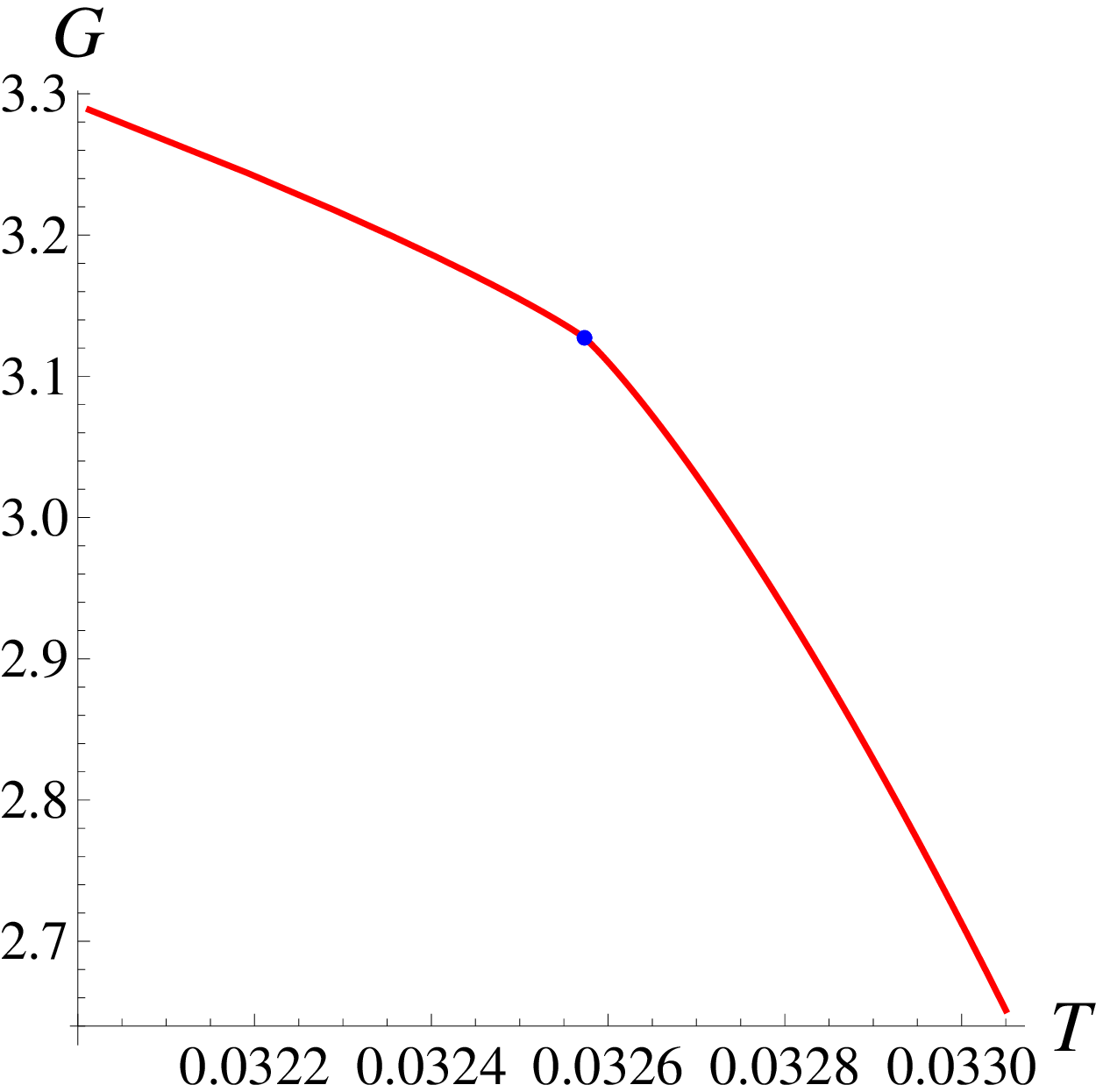}}
\caption[]{\it $G-T$ curves for RN-dS black hole with $n=5$ corresponding to
the critical effective pressure  $P_{eff}^c =0.038737$,
$P_{eff}^c =0.022365$ and $P_{eff}^c =0.012249$ respectively.
Here the Gibbs free energy
$G=M-T_{eff} S-\varphi _{eff} Q$.}}
\label{GT}
\end{figure}

Ehrenfest's equations (slope formula) for RN-dS black holes are

\be \left( {\frac{\partial P_{eff} }{\partial T_{eff} }} \right)_S
=\frac{C_P^2 -C_P^1 }{v^cT_{eff}^c (\beta _2 -\beta _1
)}=\frac{\Delta C_P }{v^cT_{eff}^c \Delta \beta }, \ee

 \be\left(
{\frac{\partial P_{eff} }{\partial T_{eff} }} \right)_v =\frac{\beta
_2 -\beta _1 }{\kappa _T^2 -\kappa _T^1 }=\frac{\Delta \beta
}{\Delta \kappa _T },\ee in which the subscript 1 and 2 represent
phase 1 and 2 respectively.

From Maxwell relation

\be \left( {\frac{\partial v}{\partial S}} \right)_P =\left(
{\frac{\partial T}{\partial P}} \right)_S \ee
and Eqs. (5.4), we have
\begin{equation}
\label{eq38}
\left[ {\left( {\frac{\partial P_{eff} }{\partial T_{eff} }} \right)_S }
\right]^c=\left[ {\left( {\frac{\partial S}{\partial v}} \right)_{P_{eff} }
} \right]^c,
\end{equation}
thus
\begin{equation}
\label{eq39}
\frac{\Delta C_P }{T_{eff}^c v^c\Delta \beta }=\left[ {\left(
{\frac{\partial S}{\partial v}} \right)_{P_{eff} } } \right]^c.
\end{equation}
Note that the footnote $`` c "$ denotes the values of physical
quantities at critical point in our letter. Here
\begin{equation}
\label{eq40}
\left( {\frac{\partial S}{\partial v}} \right)_{P_{eff} } =\left(
{\frac{\left( {\frac{\partial S}{\partial x}} \right)_{r_c } \left(
{\frac{\partial P_{eff} }{\partial r_c }} \right)_x -\left( {\frac{\partial
S}{\partial r_c }} \right)_x \left( {\frac{\partial P_{eff} }{\partial x}}
\right)_{r_c } }{\left( {\frac{\partial v}{\partial x}} \right)_{r_c }
\left( {\frac{\partial P_{eff} }{\partial r_c }} \right)_x -\left(
{\frac{\partial v}{\partial r_c }} \right)_x \left( {\frac{\partial P_{eff}
}{\partial x}} \right)_{r_c } }} \right).
\end{equation}
From Maxwell relation

\be \left( {\frac{\partial v}{\partial S}} \right)_T =\left(
{\frac{\partial T}{\partial P}} \right)_v \ee,
and Eqs. (5.5), one can get
\begin{equation}
\label{eq41}
\frac{\Delta \beta }{\Delta \kappa _T }=\left( {\frac{\partial v}{\partial
S}} \right)_{T_{eff} }^c ,
\end{equation}
here
\begin{equation}
\label{eq42}
\left( {\frac{\partial S}{\partial v}} \right)_{T_{eff} } =\left(
{\frac{\left( {\frac{\partial S}{\partial x}} \right)_{r_c } \left(
{\frac{\partial T_{eff} }{\partial r_c }} \right)_x -\left( {\frac{\partial
S}{\partial r_c }} \right)_x \left( {\frac{\partial T_{eff} }{\partial x}}
\right)_{r_c } }{\left( {\frac{\partial v}{\partial x}} \right)_{r_c }
\left( {\frac{\partial T_{eff} }{\partial r_c }} \right)_x -\left(
{\frac{\partial v}{\partial r_c }} \right)_x \left( {\frac{\partial T_{eff}
}{\partial x}} \right)_{r_c } }} \right).
\end{equation}
According to Eq. (\ref{eq34}), when $r_c \left( {\frac{\partial
T_{eff} }{\partial r_c }} \right)_x +(1-x)\left( {\frac{\partial
T_{eff} }{\partial x}} \right)_{r_c } \ne 0$, the critical points
satisfy
\begin{equation}
\label{eq43}
\left( {\frac{\partial P_{eff} }{\partial r_c }} \right)_x \left(
{\frac{\partial T_{eff} }{\partial x}} \right)_{r_c } -\left(
{\frac{\partial P_{eff} }{\partial x}} \right)_{r_c } \left( {\frac{\partial
T_{eff} }{\partial r_c }} \right)_x =0.
\end{equation}
Substituting Eq. (\ref{eq43}) into Eq. (\ref{eq42}) and comparing with Eq. (\ref{eq40}), we
have
\begin{equation}
\label{eq44}
\left( {\frac{\partial S}{\partial v}} \right)_{T_{eff} }^c =\left(
{\frac{\partial S}{\partial v}} \right)_{P_{eff} }^c .
\end{equation}
So far, we have proved that both the Ehrenfest equations are correct at the
critical point. Utilizing Eq .(\ref{eq44}), the Prigogine--Defay (PD) ratio ($\Pi
)$can be calculated as
\begin{equation}
\label{eq45}
\Pi =\frac{\Delta C_p \Delta \kappa _T }{T_{eff}^c v^c(\Delta \beta )^2}=1.
\end{equation}
Hence the phase transition occurring at $T_{eff} =T_{eff}^c $ is a second
order equilibrium transition. This is true in spite of the
fact that the phase transition curves are smeared and divergent near the
critical point. This result is in agreement with the one in\cite{Banerjee,LWB,Banerjee1,Banerjee3,Banerjee4,Banerjee5} about
AdS black holes.

\section{Conclusions}

After introducing the connection between the thermodynamic
quantities corresponding to the black hole horizon and the
cosmological horizon, we give the effective thermodynamic quantities
of the higher dimensional RN-dS system, (\ref{eq19}),(\ref{eq20})
and (\ref{eq21}). When describing the higher dimensional RN-dS
system by the effective thermodynamic quantities, it exhibits a
similar phase transition to Van der Waals equation. In Sec.4 it is
shown that the position $x$ of the phase transition point in the
higher dimensional RN-dS system is irrelevant to the electric charge
of the system. This indicates that for fixed charge when the ratio
of the black hole horizon and the cosmological horizon is $x^c$, the
second-order phase transition will occur. There are some differences
between the higher dimensional RN-dS black hole and the charged AdS
black hole. According to the isothermal curves of the Van der Waals equation and charged AdS black holes
when the temperature is lower than the critical one there exist two different phases for some values of the volume, namely  coexistence region.
From Figs.2, for the higher dimensional RN-dS black hole no
two-phase region exists for any temperature. When the effective
temperature $T_{eff}
>T_{eff}^c $, $\left( {\frac{\partial P_{eff} }{\partial v}}
\right)_{T_{eff} } <0$ which satisfy the stable condition. When
$T_{eff} <T_{eff}^c $, at some intervals the isothermal curve
corresponds to $\left( {\frac{\partial P_{eff} }{\partial v}}
\right)_{T_{eff} }>0$. At this time the system is unstable. The
states with $\left( {\frac{\partial P_{eff} }{\partial v}}
\right)_{T_{eff} }>0$ turn up at small value of $v$ with $v=r_c
(1-x)$, namely at some interval of $x$ with larger values. This means that when the both
horizons approach, the system lies at a non-equilibrium state.
Therefore, for the higher dimensional RN-dS black hole the state with coincided horizons does not exist.

In Sec. 5 we analyzed the phase transition of higher dimensional RN-dS system. It shows
that at the critical point the specific heat at constant pressure,
the volume expansivity $\beta $ and the isothermal compressibility $\kappa $
of the RN-dS system exist infinite peak, while the entropy and the
Gibbs potential G are continuous. Therefore for the phase transition
of the RN-dS system no latent heat and no specific volume changes
suddenly, it belongs to the second-order phase transition.

We carry out an analytical check of Ehrenfest equations and prove
that both Ehrenfest equations are satisfied.
This result is consistent with the nature of liquid--gas phase
transition at the critical point, hence deepening the understanding
of the analogy of RN-dS spacetime and liquid-gas systems.

\begin{acknowledgments}\vskip -4mm
This work is supported by NSFC under Grant Nos.(11175109;11075098;11247261;11205097)
and the doctoral Sustentation foundation of Shanxi Datong University
(2011-B-03).
\end{acknowledgments}


\begin{thebibliography}{07}

\bibitem{JDB1}J. D. Bekenstein,  Lett. Nuovo Cimento 4, 737 (1972).

\bibitem{JDB2}J. D. Bekenstein, Phys. Rev. D 7, 949 (1973).

\bibitem{JDB3}J. D. Bekenstein, Phys. Rev. D 9, 3292 (1974).

\bibitem{BCH}J. M. Bardeen, B. Carter, S. W. Hawking, \CMP 31, 161 (1973).

\bibitem{Hawking1}S. W. Hawking, Nature, 248, 30 (1974).

\bibitem{Hawking2}S. W. Hawking,  Commun. Math. Phys. 43 199 (1975).

\bibitem{Hawking3}S. Hawking, D. N. Page, Commun. Math. Phys. 87, 577 (1983).

\bibitem{Chamblin1}A. Chamblin, R. Emparan, C. Johnson, and R. Myers, Phys.Rev. D 60, 064018 (1999), [arXiv:hep-th/9902170].

\bibitem{Chamblin2}A. Chamblin, R. Emparan, C. Johnson, and R. Myers, Phys.Rev. D 60,104026 (1999) , [arXiv:hep-th/9904197].

\bibitem{Maldacena}O. Aharony, S. S. Gubser, J. M. Maldacena, H. Ooguri, and Y. Oz, Phys. Rept. 323, 183 (2000), [arXiv:hep-th/9905111].

\bibitem{Gubser}S. S. Gubser, Phys. Rev. D 78, 065034 (2008), [arXiv:0801.2977[hep-th]].

\bibitem{Hartnoll}S. A. Hartnoll, C. P. Herzog, and G. T. Horowitz, Phys. Rev. Lett. 101, 031601 (2008), [arXiv:0803.3295[hep-th]].

\bibitem{Sahay1}A. Sahay, T. Sarkar, and G. Sengupta, JHEP 1004, 118 (2010), [arXiv:1002.2538[hep-th]].

\bibitem{Sahay2}A. Sahay, T. Sarkar, and G. Sengupta, JHEP 1007, 082 (2010), [arXiv:1004.1625[hep-th]].

\bibitem{Sahay3}A. Sahay, T. Sarkar, and G. Sengupta, JHEP 1011, 125 (2010), [arXiv:1009.2236[hep-th]].

\bibitem{Banerjee}R. Banerjee, S. Ghosh, and D. Roychowdhury, Phys. Lett. B 696, 156 (2011), [arXiv:1008.2644[hep-th]].

\bibitem{Kastor}D. Kastor, S. Ray, J. Traschen, Class. Quant. Grav. 26, 95011 (2009), [arXiv:0904.2765[hep-th]]

\bibitem{Dolan1}B. P. Dolan, Class. Quant. Grav. 28, 125020 (2011), [arXiv:1008.5023[hep-th]].

\bibitem{SG}S. Gunasekaran, D. Kubiznak, R. B. Mann, JHEP, 1211, 110 (2012), [arXiv:1208.6251.[hep-th]].

\bibitem{BPDolan}B. P. Dolan, Phys. Rev. D 84, 127503 (2011), [arXiv:1109.0198.[hep-th]].

\bibitem{Cvetic}M. Cvetic, G.W. Gibbons, D. Kubiznak, C. N. Pope,  Phys. Rev. D 84, 024037 (2011), [arXiv:1012.2888[hep-th]].

\bibitem{Dolan2}B. P. Dolan, D Kastor, D. Kubiznak, R. B. Mann, J. Traschen, Phys. Rev. D 87, 104017 (2013), [arXiv:1301.5926[hep-th]].

\bibitem{NA1}N. Altamirano, D. Kubiznak, R. B. Mann and Z. Sherkatghanad, arXiv:1401.2586 [hep-th].

\bibitem{Dolan3}B. P. Dolan, arXiv:1209.1272[gr-qc].

\bibitem{DCZ}D. C. Zou, S. J. Zhang, B. Wang, Phys. Rev. D 89, 044002 (2014), [arXiv:1311.7299[hep-th]].

\bibitem{RBM}D. Kubiznak, R. B. Mann, JHEP 1207, 033 (2012), [arXiv:1205.0559[hep-th]].

\bibitem{LYX}S. W. Wei, Y. X. Liu, Phys. Rev. D 87, 044014 (2013), [arXiv:1209.1707[gr-qc]].

\bibitem{Cai1}R. G. Cai, L. M. Cao, L. Li, and R. Q. Yang, JHEP, 1309, 005 (2013), [arXiv:1306.6233[gr-qc]].

\bibitem{Liu}Shao-Wen Wei, Yu-Xiao Liu,  [arXiv:1402.2837[gr-qc]].

\bibitem{Hendi}S. H. Hendi, M. H. Vahidinia, Phys. Rev. D 88, 084045 (2013), [arXiv:1212.6128[hep-th]].

\bibitem{Spallucci}E. Spallucci, A. Smailagic, Phys. Lett. B 723, 436 (2013), [arXiv:1305.3379[hep-th]].

\bibitem{Belhaj}A. Belhaj, M. Chabab, H. E. Moumni and M. B. Sedra, arXiv:1306.2518[hep-th].

\bibitem{Zhao1}R. Zhao, H. H. Zhao, M. S. Ma and L. C. Zhang, Eur. Phys. J. C 73, 2645 (2013), [arXiv:1305.3725[gr-qc]].

\bibitem{MB}M. B. J. Poshteh, B. Mirza, Z.Sherkatghanad, Phys. Rev. D 88, 024005 (2013),[arXiv:1306.4516[gr-qc]].


\bibitem{LWB}J.-X. Mo, W.-B. Liu, Phys. Lett. B 727,336 (2013).

\bibitem{JXM}J.-X. Mo, W.-B. Liu, [arXiv:1401.0785[gr-qc]].

\bibitem{Banerjee1}R. Banerjee, S. K. Modak, S. Samanta, Phys. Rev. D84, 064024 (2011), [arXiv:1005.4832[hep-th]].

\bibitem{Banerjee3}R. Banerjee, D. Roychowdhury, JHEP 11, 004 (2011), [arXiv:1109.2433[hep-th]].
\bibitem{Banerjee4}R. Banerjee, D. Roychowdhury, Phys. Rev. D 85, 044040 (2012), [arXiv:1111.0147[hep-th]].
\bibitem{Banerjee5}R. Banerjee, D. Roychowdhury, Phys. Rev. D 85, 104043 (2012), [arXiv:1203.0118[hep-th]].

\bibitem{NA}N. Altamirano, D. Kubiznak and R. Mann, Phys. Rev. D 88, 101502 (2013), [arXiv:1306.5756 [hep-th]].

\bibitem{Ma}M. S. Ma, H. H. Zhao, L. C. Zhang, R. Zhao, \IJMPA,
to be published.

\bibitem{Cai2}R. G. Cai, \NPB 628, 375 (2002).

\bibitem{Sekiwa}Y. Sekiwa, Phys. Rev. D 73, 084009(2006),[arXiv:hep-th/0602269].

\bibitem{Urano}M. Urano, A. Tomimatsu, Class. Quant. Grav.26, 105010 (2009), [arXiv:0903.4230[hep-th]].

\bibitem{Zhang}L. C. Zhang, H. F. Li and R. Zhao, Science China, Physics, Mechanics Astronomy 54, 1384 (2011).

\bibitem{Myung}Y. S. Myung, Phys. Rev. D 77, 104007 (2008).

\bibitem{Kim}M. Eune, W. Kim, Phys. Lett. B 723, 177 (2013).

\bibitem{Cai3}R. G. Cai, Physics 34, 555(2005),( in Chinese).

\bibitem{SB}S. Bhattacharya, A. Lahir, Eur. Phys. J. C 73, 2673 (2013), [arXiv:1301.4532[gr-qc]].

\bibitem{Cai4}R. G. Cai, J. Y. Ji, K. S. Soh, Classical. Quantum. Grav. 15, 2783 (1998).

\bibitem{realroots}B. D. Koberlein and R. L. Mallett, Phys. Rev. D 49,
5111(1994).

\bibitem{Gibbons1}G. W. Gibbons, H. L\"{u }, D. N. Page, and C. N. Pope, Phys. Rev. Lett. 93, 171102 (2004), [arXiv:hep-th/0409155].

\bibitem{Gibbons2}G. W. Gibbons, H. L\"{u }, D. N. Page, and C. N. Pope, J. Geom. Phys. 53, 49(2005), [arXiv:hep-th/0404008].

\bibitem{Zhao2}R. Zhao, L. C. Zhang, H. F. Li,  Gen. Relat. Grav. 42, 975 (2010).

\bibitem{Zhao3}R. Zhao, L. C. Zhang, H. F. Li and Y. Q. Wu,  Eur. Phys. J. C. 65, 289 (2010).


\end{thebibliography}
\end{document}